\documentclass[showpacs,aps,prd,nofootinbib,floatfix,amsmath,amssymb,twocolumn]{revtex4}
\usepackage{mathrsfs}
\usepackage{graphicx}
\usepackage[dvipsnames]{xcolor}
\usepackage{dsfont}
\usepackage{hyperref}
\usepackage[noabbrev]{cleveref}
\hypersetup{colorlinks=true, linkcolor=blue, citecolor=blue}
\usepackage{orcidlink}
\begin{document}

\makeatletter
\newbox\slashbox \setbox\slashbox=\hbox{$/$}
\newbox\Slashbox \setbox\Slashbox=\hbox{\large$/$}
\def\pFMslash#1{\setbox\@tempboxa=\hbox{$#1$}
  \@tempdima=0.5\wd\slashbox \advance\@tempdima 0.5\wd\@tempboxa
  \copy\slashbox \kern-\@tempdima \box\@tempboxa}
\def\pFMSlash#1{\setbox\@tempboxa=\hbox{$#1$}
  \@tempdima=0.5\wd\Slashbox \advance\@tempdima 0.5\wd\@tempboxa
  \copy\Slashbox \kern-\@tempdima \box\@tempboxa}
\def\FMslash{\protect\pFMslash}
\def\FMSlash{\protect\pFMSlash}
\def\miss#1{\ifmmode{/\mkern-11mu #1}\else{${/\mkern-11mu #1}$}\fi}
\makeatother

\title{
$WWh$ anomalous couplings at one loop from a seesaw variant of radiatively-induced neutrino masses}

\author{H\'ector Novales-S\'anchez\orcidlink{0000-0003-1503-3086}$^{(a)}$}
\author{Enrique Ram\'irez\orcidlink{0000-0002-8635-4621}$^{(a)}$}
\author{M\'onica Salinas\orcidlink{0000-0001-6413-9552}$^{(b)}$}
\author{Humberto V\'azquez-Castro\orcidlink{0009-0006-7830-9114}$^{(a)}$}
\affiliation{
$^{(a)}$Facultad de Ciencias F\'isico Matem\'aticas, Benem\'erita Universidad Aut\'onoma de Puebla, Apartado Postal 1152 Puebla, Puebla, M\'exico\\$^{(b)}$Departamento de F\'isica, Centro de Investigaci\'on y de Estudios Avanzados del IPN, Apartado Postal 14-740, 07000 Ciudad de M\'exico, M\'exico}

\begin{abstract}
After the successful measurement of the Higgs-like particle at the Large Hadron Collider, the determination of whether it corresponds to either the minimal version of the Higgs scalar, provided by the Standard Model, or to some new physics, so far unknown, has become a main objective of the high-energy-physics community. In particular, the investigation of the couplings of this particle with the rest of the Standard-Model field content is a central task. To this aim, in the present paper we consider a variant of the seesaw neutrino-mass-generating mechanism in which the masses of light neutrinos are generated via radiative corrections of the neutrino 2-point function. In this framework, we calculate the contributions from virtual Majorana neutrinos to the $WWh$ vertex, at one loop. A set of anomalous couplings, some of them $CP$-even and the others $CP$-odd, emerge from this calculation. We perform numerical estimations of these anomalous couplings by taking into account the role of the $WWh$ vertex in (1) $Wh$ production from $pp$ collisions and (2) in Higgs production by vector-boson fusion, occurring in some lepton collider. Our estimations show that all the generated anomalous couplings are beyond current experimental sensitivity. However, we note that $CP$-conserving anomalous couplings could be as large as $\sim10^{-3}$, which would be near the reach of the high-luminosity phase of the Large Hadron Collider. On the other hand, $CP$-odd effects, amounting to $\sim10^{-4}$, turn out to be well below expectations for the Compact Linear Collider by about 2 orders of magnitude. 
\end{abstract}

\pacs{}

\maketitle

\section{Introduction}
\label{theintro}
Nowadays, the Standard Model~\cite{Glashow:1961tr,Salam:1968rm,Weinberg:1967tq} (SM) still remains the best description ever achieved of fundamental physics, not only by virtue of its elegant mathematical structure, supported by the principle of symmetry, but also due to its so far remarkable consistency with data provided by the most powerful experimental facilities ever built~\cite{ParticleDataGroup:2024cfk}. The measurement, nearly fifty years after the theoretical realization of the Brout-Englert-Higgs mechanism~\cite{Higgs:1964pj,Englert:1964et}, of a scalar boson with mass $\approx 125\,{\rm GeV}$ has been a major event~\cite{ATLAS:2012yve,CMS:2012qbp}, not only for the high-energy physics community, but for the entire world. This discovery has shown that the origin of mass is the spontaneous breaking of gauge symmetry, a key ingredient for the construction of sensible and realistic field theories. While so far this scalar particle looks a lot like the SM version of the Higgs boson, the possibility that it rather corresponds to some physical formulation other than the SM must be recognized and thoroughly explored. Getting to a final conclusion is a quite difficult task, which will expectedly take several years, as it requires a great amount of theoretical, experimental, and phenomenological studies. Therefore, the exploration and analyses of the couplings of the Higgs boson, together with their measurements at high precision level, have become a priority in the scientific agenda for the forthcoming decades. 
\\

The occurrence of Higgs-boson couplings other than those which characterize the SM offers interesting possibilities, bearing great relevance, for deviations from SM predictions are to be interpreted as manifestations of unknown physics. For the present work, we have devoted our attention to the vertex $WWh$, which characterizes how two SM $W$ gauge bosons and a Higgs scalar boson, $h$, couple. The well-known tree-level Lorentz-covariant structure of this vertex, in the SM framework, is 
\begin{equation}
\begin{gathered}
\vspace{-0.15cm}
\includegraphics[width=2.5cm]{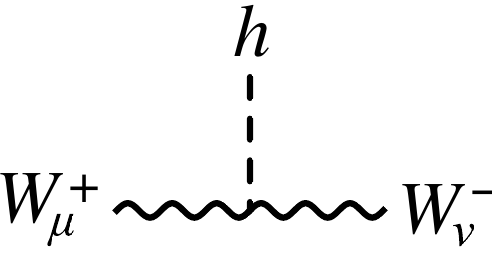}
\end{gathered}=ig\,m_W g_{\mu\nu},
\end{equation}
with $g$ the ${\rm SU}(2)_L$ coupling constant and $m_W$ the mass of the $W$ boson. The effective-Lagrangian technique, a model-independent mean to estimate at low, currently accessible, energies the effects from the genuine fundamental high-energy theory beyond the SM, opens the possibility of having a richer set of $WWh$ couplings~\cite{Giudice:2007fh,Grzadkowski:2010es,LHCHiggsCrossSectionWorkingGroup:2016ypw,Dutta:2008bh}, which can even include $CP$-odd effects. A bunch of works have addressed $WWh$ anomalous couplings (AC) from the perspective of the effective-Lagrangian approach, aiming at estimations of the plausibility of measuring effects from new physics through diverse physical processes, to take place in hadron~\cite{Ellis:2014dva,Nakamura:2017ihk,Alloul:2013naa,Fabbrichesi:2023jep,Mimasu:2015nqa,Bizon:2021rww}, lepton-hadron~\cite{Kumar:2015kca,Sharma:2022epc}, and lepton colliders~\cite{Dutta:2008bh,ILC:2013jhg,Denizli:2017pyu}. The heretofore observed consistency of experimental data with SM predictions is an aspect which has strongly motivated the usage of effective Lagrangians to address possible further Higgs couplings, instead of exploring the impact of specific SM extensions. As pointed out by the author of Ref.~\cite{Peskin:2022pfv}, the expected experimental sensitivity for the next years in comparison with the expected impact from new physics on Higgs-boson couplings renders precision Higgs physics a distant objective, at least for now. However, the very same author investigates SM extensions after claiming that new physics frameworks producing sizable effects on Higgs couplings may exist. With this in mind, we have explored the one-loop contributions from an extension of the SM, featuring massive neutrinos, to the $WWh$ vertex. 
\\

Even though agreement of experimental data and SM predictions seems to be the rule, the existence of new physics, associated to phenomena out of the reach of the SM, is factual indeed. Issues clearly supported by experimental data, such as the problem of dark matter~\cite{Zwicky:1933gu,Rubin:1970zza,DES:2015ttt}, lack an explanation from the SM. Another instance of the SM limitations lies in its neutrino sector, which has been proven to be, at best, a good approximation, since neutrinos are assumed to be massless in this model, which contradicts the confirmed occurrence of neutrino oscillations~\cite{Super-Kamiokande:1998kpq,SNO:2002tuh,DayaBay:2012fng,RENO:2012mkc}, an outcome of neutrino non-zero masses and mixing~\cite{Pontecorvo:1957cp}. The SM neutrino sector, now understood to be incomplete, is a well-spotted place where new physics resides for sure, and therefore has received a great deal of attention for the last 25 years or so. In particular, the massiveness and electric neutrality of neutrinos has come along with the exciting possibility that the proper characterization of these particles corresponds not to Dirac fermion fields~\cite{Dirac:1928hu}, but to the Majorana description~\cite{Majorana:1937vz}, in which case the neutrino fields, $\nu_j$, match their charge-conjugate fields, $\nu_j^{\rm c}=C\overline{\nu_j}^{\rm T}$, with $C$ the charge-conjugation matrix. A confirmation of neutrinos described {\it alla} Majorana would be a milestone, not only by dint of the information of the new physics this would come along with, but also because no Majorana elementary particle has ever been measured and no one knows for certain whether such a sort of particles actually exists in nature. And speaking of neutrino massiveness, a well-known mean to define masses for these particles is the seesaw mechanism, born from an attempt to explain parity violation as a spontaneously-broken symmetry~\cite{Mohapatra:1979ia}. A quite representative feature of the seesaw mechanism is the hypothetical existence of heavy partners, $N_j$, frequently dubbed ``heavy neutral leptons'' or ``heavy neutrinos'', characterized by both electric neutrality and, of course, large masses. In this sense, the already known neutrinos are sometimes referred to as ``light neutrinos''. Being $\Lambda$ the high-energy scale associated to some stage of spontaneous symmetry breaking, at which Majorana-mass-like terms are generated, light-neutrino masses turn out to be given by $m_\nu\sim\frac{v^2}{\Lambda}$, with $v=246\,{\rm GeV}$ the vacuum expectation value of the Higgs potential, whereas the masses of heavy neutrinos are $m_N\sim\Lambda$. In this manner, the link $m_{N}\sim\frac{v^2}{m_{\nu}}$, among the masses of the two types of neutrinos, is defined. A number of variants of this mechanism\footnote{Ref.~\cite{Cai:2017mow} provides a nice review on seesaw variants.}, each bearing its very own assumptions, have been realized, among which the inverse seesaw~\cite{Mohapatra:1986bd,Gonzalez-Garcia:1988okv,Deppisch:2004fa} and the linear seesaw~\cite{Akhmedov:1995ip,Akhmedov:1995vm} have been widely studied. The motivation for the realization of such seesaw-like mechanisms of neutrino-mass generation has been mainly fed by the restriction imposed by current bounds on light-neutrino masses~\cite{eBOSS:2020yzd,Planck:2018vyg,CUORE:2019yfd,GERDA:2020xhi,KamLAND-Zen:2016pfg,KATRIN:2024cdt}, which, in the original versions of the seesaw mechanism, enforce the scale $\Lambda$ to be enormous, then making both direct production and virtual effects of heavy neutrinos well out of the reach of experimental facilities at present and even in the long term. 
\\

For the present investigation, we have considered the seesaw variant devised by the author of Ref.~\cite{Pilaftsis:1991ug}, in which the masses of light neutrinos vanish at the tree level. Then, light neutrinos become massive through radiative corrections of the neutrino 2-point function. Unlike what happens in the inverse and the linear low-scale seesaw variants, in which the introduction of a further low-energy scale weakens the neutrino-mass seesaw relation $m_{N}\sim\frac{v^2}{m_{\nu}}$, the aforementioned elimination of tree-level light-neutrino masses indeed cancels such a relation, while a new link among the masses of light and heavy neutrinos, driven by the radiative corrections which generate the masses of the former, emerges: the mass spectrum of heavy neutrinos must be quasi-degenerate in order to define small radiative masses for the light neutrinos. The one-loop contributions to $WWh$ follow from the consideration of charged currents featuring the SM $W$-boson field and from the couplings of neutrinos to the $Z$ and Higgs bosons. The corresponding Lagrangians are
\begin{eqnarray}
&&
{\cal L}_{W\nu l}=\sum_{\alpha=e,\mu,\tau}\sum_{j=1}^3\frac{g}{\sqrt{2}}
\big(
{\cal B}_{\alpha\nu_j}W^-_\rho\overline{l_\alpha}\gamma^\rho P_L\nu_j
\nonumber \\ && \hspace{2cm}
+{\cal B}_{\alpha N_j}W^-_\rho\overline{l_\alpha}\gamma^\rho P_LN_j
\big)+{\rm H. c.},
\label{LWnl}
\end{eqnarray}
\begin{eqnarray}
&&
{\cal L}_{h\nu\nu}=\sum_{k=1}^3\sum_{j=1}^3
\frac{g}{4m_W}h\Big(
\overline{\nu_k}
\big(
(m_{\nu_k}+m_{\nu_j}){\cal C}^{\rm Re}_{\nu_k\nu_j}
\nonumber \\
&&
-i\gamma_5(m_{\nu_k}-m_{\nu_j}){\cal C}^{\rm Im}_{\nu_k\nu_j}
\big)\nu_j
+2\overline{\nu_k}
\big(
(m_{\nu_k}+m_{N_j}){\cal C}^{\rm Re}_{\nu_k N_j}
\nonumber \\ 
&&
-i\gamma_5(m_{\nu_k}-m_{N_j}){\cal C}^{\rm Im}_{\nu_kN_j}
\big)N_j
+\overline{N_k}
\big(
(m_{N_k}+m_{N_j}){\cal C}^{\rm Re}_{N_kN_j}
\nonumber \\ &&
-i\gamma_5(m_{N_k}-m_{N_j}){\cal C}^{\rm Im}_{N_kN_j}
\big)N_j
\Big).
\label{Lhnn}
\end{eqnarray}
\begin{eqnarray}
&&
{\cal L}_{Z\nu\nu}=\sum_{k=1}^3\sum_{j=1}^3
\frac{g}{4c_{\rm W}}\Big(
Z_\rho\overline{\nu_k}\gamma^\rho\big( i{\cal C}^{\rm Im}_{\nu_k\nu_j}-{\cal C}^{\rm Re}_{\nu_k\nu_j}\gamma_5 \big)\nu_j
\nonumber \\ &&
\hspace{0.8cm}
+\Big( Z_\rho\overline{\nu_k}\gamma^\rho\big( i{\cal C}^{\rm Im}_{\nu_kN_j}-{\cal C}^{\rm Re}_{\nu_kN_j}\gamma_5 \big)N_j+{\rm H.c.} \Big)
\nonumber \\ &&
\hspace{0.8cm}
+Z_\rho\overline{N_k}\gamma^\rho\big( i{\cal C}^{\rm Im}_{N_kN_j}-{\cal C}^{\rm Re}_{N_kN_j}\gamma_5 \big)N_j
\Big),
\label{LNC}
\end{eqnarray}
where $\nu_j$ and $N_j$, with $j=1,2,3$, denote the light- and heavy-neutrino fields, respectively. The masses of light neutrinos have been denoted by $m_{\nu_j}$, whereas heavy-neutrino masses are represented by $m_{N_j}$. Note that, since tree-level light-neutrino masses vanish, in accordance with the model, $m_{\nu_j}=0$ has to be taken in Eq.~\eqref{Lhnn}. Spinor fields corresponding to charged leptons are represented by $l_\alpha$, with $\alpha=e,\mu,\tau$. In addition, $W^+_\mu$ and $W^-_\mu$ are $W$-boson fields, $Z_\mu$ is the $Z$-boson field, and $h$ is the Higgs-boson field. The short-hand notation $c_{\rm W}=\cos\theta_{\rm W}$, where $\theta_{\rm W}$ is the weak-mixing angle, has been used as well. Eq.~\eqref{LWnl} involves the factors ${\cal B}_{\alpha\nu_j}$ and ${\cal B}_{\alpha N_j}$, which are the entries of two $3\times3$ matrices, respectively denoted by ${\cal B}_\nu$ and ${\cal B}_N$, constituting the $3\times6$ matrix 
\begin{equation}
{\cal B}=
\left( 
\begin{array}{cc}
{\cal B}_\nu & {\cal B}_N
\end{array}
\right).
\label{Bmatrix}
\end{equation} 
Such a matrix fulfills a sort of semi-unitarity condition, namely, ${\cal B}{\cal B}^\dag={\bf 1}_3$, with ${\bf 1}_3$ denoting the $3\times3$ identity matrix. A $6\times6$ Hermitian matrix, ${\cal C}$, is given, on the other hand, by the product ${\cal B}^\dag{\cal B}={\cal C}$, and is conveniently expressed, in terms of $3\times3$ matrix blocks, as
\begin{equation}
{\cal C}=
\left(
\begin{array}{cc}
{\cal C}_{\nu\nu} & {\cal C}_{\nu N}
\vspace{0.3cm}
\\
{\cal C}_{N\nu} & {\cal C}_{NN}
\end{array}
\right).
\label{Cmatrix}
\end{equation}
Note that ${\cal C}={\rm Re}\big\{ {\cal C} \big\}+i\,{\rm Im}\big\{ {\cal C} \big\}$, so, in Eqs.~(\ref{Lhnn}) and (\ref{LNC}), we have denoted ${\cal C}^{\rm Re}={\rm Re}\big\{ {\cal C} \big\}$ and ${\cal C}^{\rm Im}={\rm Im}\big\{{\cal C}\big\}$. This matrix also fulfills ${\cal C}{\cal C}^\dag={\cal C}$. 
\\\\\\\\

In the aforedescribed framework, we calculate the contributions, at one loop, to $WWh$, produced by Eqs.~(\ref{LWnl}) to (\ref{LNC}), and then identify form factors characterizing the Lorentz covariant parametrization of this vertex. The calculation is carried out by using the tensor-reduction method~\cite{Passarino:1978jh,Devaraj:1997es}, along with usage of dimensional regularization~\cite{Bollini:1972ui,tHooft:1972tcz}, aimed at dealing with latent ultraviolet (UV) divergences encoded within 4-momentum integrals. All the contributions, but the one matching the SM coupling, are found to be finite in the UV sense. From the one-loop calculation, a set of 4 nonzero independent $WWh$ AC emerges. Since the calculation is executed off shell, all the AC depend on squared momenta of the external lines. To deal with this, we contextualize our numerical estimations of these quantities by considering two cases: one $W$-boson line off shell, whereas the other $W$-boson line and the Higgs-boson line are on shell; the two $W$-boson lines are assumed to be off shell, but the Higgs-boson line is taken on shell. The aforementioned contextualization for these two cases consists in considering physical processes in which the $WWh$ vertex plays a role. One of such processes is the production of pairs $Wh$ from proton-proton collisions, whereas the other is vector-boson fusion (VBF) in lepton colliders, that is $\ell\ell\to\nu\nu h$. Our estimation yield $CP$-preserving contributions as large as $\sim10^{-3}$, almost within experimental sensitivity of the high-luminosity Large Hadron Collider (HL-LHC). On the other hand, $CP$-violating effects lie, of order $10^{-4}$, well below projected sensitivity for the Compact Linear Collider (CLIC).
\\

The rest of the paper develops in the following manner: Section~\ref{thecalc} is dedicated to the description of the analytical calculation, including that of the one-loop Feynman diagrams and the proper Lorentz-covariant parametrization of the $WWh$ vertex; in Section~\ref{thenums} we present our estimations and analysis of the generated contributions to the AC; finally, a summary and our conclusions are given in Section~\ref{theend}.


\section{The \texorpdfstring{$WWh$}\text{\hspace{2mm}vertex and the new physics}}
\label{thecalc}
Higgs physics is nowadays very active, with a whole lot of work to be done and a plethora of investigations being carried out right now. Since the measurement, by the ATLAS and CMS Collaborations, of the scalar Higgs-like particle, the determination of whether such a finding actually corresponds to the SM Higgs boson has become a main objective of the high-energy physics research agenda. This program includes, as a central goal, the detailed examination of the Higgs couplings with all other SM particles, aiming at the elucidation of the plausibility that these couplings carry effects from some unknown underlying physics, which would presumably manifest as slight deviations from SM predictions. Moreover, physical descriptions beyond the SM might come along with Higgs couplings characterized by Lorentz structures other than those found in the SM at the tree level, or even at the loop level. In this context, the gauge couplings of the Higgs boson serve as probes of the gauge structure of the SM. With this in mind, we have considered, for the present investigation, the $WWh$ vertex, which characterizes the couplings of two SM $W$'s and a Higgs boson. This vertex participates in physical processes of interest. In the SM framework, the process $h\to WW^*$ turns out to be among the Higgs-boson preferred decay modes~\cite{ParticleDataGroup:2024cfk,LHCHiggsCrossSectionWorkingGroup:2011wcg}. The one-loop order radiative corrections to this decay process, in the SM, have been calculated in Ref.~\cite{Kniehl:1991xe}. Another remarkable process involving the vertex $WWh$ is vector boson fusion $WW\to h$, as it is one of the main Higgs production mechanisms in both hadron colliders~\cite{LHCHiggsCrossSectionWorkingGroup:2016ypw}, just behind gluon fusion, and electron-positron colliders~\cite{ILC:2013jhg}. There is also the Higgs decay into four fermions, $h\to WW\to4f$, which has drawn some attention. The authors of Ref.~\cite{Bredenstein:2006rh} carried out an intricate and comprehensive calculation of the decay rate for this process, at the one-loop level, in the context of the SM. The CMS Collaboration reported, in Ref.~\cite{CMS-PAS-HIG-13-005}, a measurement of the Higgs-boson mass, achieved through combined investigations on several processes, including $h\to WW^*\to2\ell2\nu$. Gluon fusion and VBF were measured by the ATLAS Collaboration by studying the decay $h\to WW^*\to e\nu\mu\nu$~\cite{ATLAS:2022ooq}. A recent investigation by the CMS Collaboration, Ref.~\cite{CMS:2024srp}, considered a rare process characterized by pairs $\overline{b}b$ and $\ell\nu$ in the final state, along with two jets. Such a process, in which $\overline{b}b$ follows from Higgs production, involves both gauge-Higgs couplings $WWh$ and $ZZh$. The process $ff\to\nu\overline{\nu}h$ is another notorious process, in which Higgs production follows from VBF $WW\to h$. This sort of processes is expected to play a relevant role for precision studies at electron-positron colliders. According to The International Linear Collider Technical Design Report~\cite{ILC:2013jhg}, Higgs production from VBF $WW\to h$ will provide high-precision measurements of the $WWh$ coupling from $e^+e^-$ collisions at a center-of-mass energy (CME) of $500\,{\rm GeV}$, which, furthermore, will make it possible to investigate the Lorentz structure of such an interaction.
\\

\subsection{Anomalous \texorpdfstring{$WWh$}\text{\hspace{2mm}couplings}}
While several models extending the SM are available, with a large diversity of assumptions regarding the new physics, the formalism of effective Lagrangians, distinguished from SM extensions for being independent of what the genuine high-energy formulation actually looks like, offers an alternative approach to estimate and analyze the effects, at energies within the reach of current experimental facilities, from unknown high-energy physics. Once the low-energy dynamic variables and symmetries are given, the whole set of effective-Lagrangian terms are determined~\cite{Wudka:1994ny}. In particular, the effective Lagrangian for the SM has the general form
\begin{equation}
{\cal L}^{\rm eff}_{\rm SM}={\cal L}_{\rm SM}+\sum_{n>4}^\infty\sum_{j=1}^{j_n}\frac{\alpha_{j}^{(n)}}{\Lambda^{n-4}}{\cal O}_{j}^{(n)}.
\end{equation}
In this expression, ${\cal L}_{\rm SM}$ is the SM renormalizable Lagrangian, followed by an infinite series of non-renormalizable terms generated by integrating out the heavy fields associated to the high-energy theory~\cite{Dobado:1997jx}. Then $\Lambda$ is to be understood as the energy scale at which the fundamental description operates in full, whereas the coefficients $\alpha_j^{(n)}$ are dimensionless quantities, called ``Wilson coefficients'', determined by parameters of the high-energy theory. Further, each factor ${\cal O}^{(n)}_j$ stands for a product of SM fields and derivatives, with units $({\rm mass})^{n}$ and with its structure determined by the symmetries of the low-energy theory, in this case those of the SM. In particular, ${\cal L}_{\rm SM}^{\rm eff}$ is governed by the gauge-symmetry group ${\rm SU}(3)_C\otimes{\rm SU}(2)_L\otimes{\rm U}(1)_Y$. Therefore, the factor $\frac{1}{\Lambda^{n-4}}$ introduces a suppression on the effects from the new physics, at the same time that it amends the units of all the non-renormalizable terms ${\cal O}^{(n)}_j$, for each $n>4$, in such a way that ${\cal L}^{\rm eff}_{\rm SM}$ has the correct units $({\rm mass})^4$. If lepton number is preserved, then the allowed non-renormalizable terms correspond exclusively to $n=6,8,10,\ldots$, whereas the assumption of lepton-number violation loosens restrictions, thus allowing the occurrence of terms associated to $n=5,7,9,\ldots$~\cite{Babu:2001ex}, of which the very first in the list is the Weinberg operator~\cite{Weinberg:1979sa}. While the number of terms in any effective Lagrangian is, in principle, infinite, thus implying the presence of an infinite number of parameters, the assumption of a large energy scale $\Lambda$ defines a clear hierarchy in the sense that the larger the $n$ the larger the $\frac{1}{\Lambda^{n-4}}$ suppression. In practical terms, this translates into the possibility of disregarding most non-renormalizable terms, only leaving a finite subset comprised by the terms with canonical dimension $({\rm mass})^n$ up to some desired $n$. This has motivated the exploration of the set of effective-Lagrangian terms distinguished by having units $({\rm mass})^6$~\cite{Leung:1984ni,Buchmuller:1985jz}, since the suppression introduced by $\Lambda$ is the smallest among all the non-renormalizable terms and because such terms introduce a rich phenomenology.
\\

Among the whole set of ${\cal L}_{\rm SM}^{\rm eff}$ Lagrangian terms with canonical dimension 6, some generate $WWh$ couplings, once electroweak symmetry breaking has taken place. A large number of such terms can be constructed on the grounds of the criteria provided by symmetry, but not all of them are independent of each other~\cite{Buchmuller:1985jz,Wudka:1994ny,Grosse-Knetter:1993tae}. This observation has given rise to several bases of Higgs-associated effective-Lagrangian terms. One is the so-called SILH basis, the acronym for Strongly-Interacting Light Higgs, which was originally proposed, in Ref.~\cite{Giudice:2007fh}, with the objective of characterizing the phenomenology associated to scenarios of a light Higgs boson with strong dynamics. The SILH Lagrangian is given as
\begin{eqnarray}
&&
{\cal L}_{\rm SILH}=\frac{\overline{c}_\Phi}{2v^2}\partial^\mu\big( \Phi^\dag\Phi \big)\partial_\mu\big( \Phi^\dag\Phi \big)
\nonumber \\ &&
\hspace{1cm}
+\frac{ig\overline{c}_{\Phi W}}{m_W^2}\big( (D^\mu\Phi)^\dag\sigma^jD^\nu\Phi \big)W^j_{\mu\nu}
\nonumber \\ &&
\hspace{1cm}
+\frac{ig\overline{c}_W}{2m_W^2}\big(\Phi^\dag \sigma^j \overset{\text{\scriptsize$\leftrightarrow$}}{D}\hspace{0.0000000001cm}^\mu\Phi \big) D^\nu W^j_{\mu\nu}
\nonumber \\ &&
\hspace{1cm}
+\frac{ig\tilde{c}_{\Phi W}}{m_W^2}\big( (D^\mu\Phi)^\dag\sigma^jD^\nu\Phi \big)\tilde{W}^j_{\mu\nu}
+\cdots,
\label{LSILH}
\end{eqnarray}
for which use has been made of the conventions and notation of Ref.~\cite{Alloul:2013naa}. Only four terms of ${\cal L}_{\rm SILH}$ are shown explicitly in the last equation, while the ellipsis represents further terms, not to be taken into account for the present investigation. Moreover, $\Phi$ is the Higgs doublet, $D_\mu$ is the ${\rm SU}(2)_L\otimes{\rm U}(1)_Y$ covariant derivative, and $W^j_{\mu\nu}=\partial_\mu W^j_\nu-\partial_\nu W^j_\mu+g\,\epsilon^{jki}W^k_\mu W^i_\nu$ are the Yang-Mills field strengths associated to ${\rm SU}(2)_L$ gauge symmetry. For Eq.~(\ref{LSILH}) to be written down, $g\,\overset{\text{\scriptsize$\leftrightarrow$}}{D}\hspace{0.0000000001cm}^\mu f=g\,\overset{\text{\tiny$\rightarrow$}}{D}\hspace{0.0000000001cm}^\mu f-g\,\overset{\text{\tiny$\leftarrow$}}{D}\hspace{0.0000000001cm}^\mu f$ and the dual field strength $\tilde{W}^j_{\mu\nu}=\frac{1}{2}\epsilon_{\mu\nu\rho\sigma}W^{j\rho\sigma}$ have been defined. We also have the Wilson coefficients $\overline{c}_\Phi$, $\overline{c}_{\Phi W}$, $\overline{c}_W$, and $\tilde{c}_{\Phi W}$. After implementation of the Higgs mechanism to ${\cal L}_{\rm SILH}$, Eq.~(\ref{LSILH}), the resulting $WWh$ couplings, governed by the electromagnetic gauge group ${\rm U}(1)_e$, read
\begin{eqnarray}
&&
{\cal L}_{\rm SILH}=g \,m_W\lambda_1 W^+_\mu W^{-\mu}h
-\frac{g\lambda_2}{2m_W}W^{+\mu\nu}W^-_{\mu\nu} h
\nonumber \\ &&
\hspace{1cm}
-\frac{g\lambda_3}{m_W}\big( W^{+\nu}\partial^\mu W^-_{\mu\nu}+{\rm H.c.} \big)h
\nonumber \\ &&
\hspace{1cm}
-\frac{g\tilde{\lambda}_1}{2m_W}W^{+\mu\nu}\tilde{W}^-_{\mu\nu}h+\cdots,
\label{LSILHassb}
\end{eqnarray}
with $\lambda_1=1-\frac{\overline{c}_\Phi }{2}$, $\lambda_2=2\,\overline{c}_{\Phi W}$, $\lambda_3=\overline{c}_{HW}+\overline{c}_W$, and $\tilde{\lambda}_1=2\,\tilde{c}_{\Phi W}$. Such effective-Lagrangian terms have been considered in a number of papers, for instance in Refs.~\cite{Maltoni:2013sma,Ellis:2014dva,Denizli:2017pyu,Kumar:2015kca,Nakamura:2017ihk,Alloul:2013naa,Sharma:2022epc,Yan:2021tmw,Fabbrichesi:2023jep,Mimasu:2015nqa,Bizon:2021rww,Dutta:2008bh,Han:2005pu}, to investigate the impact of presumable new physics on $WWh$, by addressing a variety of processes in which this vertex participates. Now consider an effective Lagrangian, ${\cal L}_{WWh}^{\rm eff}$, which comprises the terms explicitly displayed in Eq.~(\ref{LSILHassb}) and further terms. Such a Lagrangian, invariant under the electromagnetic group, is expressed as the sum ${\cal L}_{WWh}^{\rm eff}={\cal L}_{WWh}^{\rm even}+{\cal L}_{WWh}^{\rm odd}$, with ${\cal L}_{WWh}^{\rm even}$ invariant with respect to the discrete $CP$ transformation and where ${\cal L}_{WWh}^{\rm odd}$ is characterized by $CP$ violation. These Lagrangian terms, distinguished by their definite $CP$ properties, are given as
\begin{eqnarray}
&&
{\cal L}_{WWh}^{\rm even}=g \,m_W\lambda_1 W^+_\mu W^{-\mu}h
\nonumber \\ &&
\hspace{0.5cm}
-\frac{g}{m_W}\Big(
\frac{\lambda_2}{2}W^{+\mu\nu}W^-_{\mu\nu}
+\lambda_3
\big( W^{+\nu}\partial^\mu W^-_{\mu\nu}+{\rm H.c.} \big)
\Big)h
\nonumber \\ &&
\hspace{1cm}
-\frac{g}{m_W^3}\Big(
\lambda_4\big( W^{+\nu}\partial_\mu W_{\nu\rho}^-+{\rm H.c.} \big)
\nonumber \\ &&
\hspace{1.5cm}
+\lambda_5W^{+\nu}\hspace{0.00001cm}_{\rho}W^-_{\nu\mu}
\Big)\partial^\mu\partial^\rho h,
\label{LCPeven}
\end{eqnarray}
\begin{eqnarray}
&&
{\cal L}_{WWh}^{\rm odd}=-\frac{g\tilde{\lambda}_1}{2m_W}W^{+\mu\nu}\tilde{W}^-_{\mu\nu}h
\nonumber \\ &&
\hspace{1.5cm}
-\frac{g\tilde{\lambda}_2}{m_W}\big( i\,W^{+\nu}\partial^\mu W^-_{\mu\nu}+{\rm H.c.} \big)h,
\label{LCPodd}
\end{eqnarray}
where the $W$-boson field $W_\mu$ defines the 2-tensors $W^+_{\mu\nu}=\partial_\mu W^+_\nu-\partial_\nu W^+_\mu$ and $W^-_{\mu\nu}=\partial_\mu W^-_\nu-\partial_\nu W^-_\mu$. About the $CP$-even Lagrangian ${\cal L}^{\rm even}_{WWh}$, Eq.~(\ref{LCPeven}), note that the terms involving $\lambda_4$ and $\lambda_5$ are not generated by mass-dimension-6 terms, while they receive contributions from Lagrangian terms, invariant under ${\rm SU}(2)_L\otimes{\rm U}(1)_Y$, with canonical dimension $\geqslant8$. This can be seen from the number of derivatives in each of the associated couplings. For instance, consider the following effective-Lagrangian terms:
\begin{equation}
{\cal L}_{D\Phi}=\frac{g\,\overline{c}_{D\Phi}}{2m_W^4}\big(\Phi^\dag D^\mu D^\rho\Phi\big)\,{\rm tr}\big\{ W^\nu\hspace{0.000001cm}_\rho W_{\nu\mu} \big\}+{\rm H.c.},
\label{LPhibsb}
\end{equation}
\begin{equation}
{\cal L}_{DW}=\frac{\overline{c}_{DW}}{2m_W^4}
i\big(( D^\nu\Phi )^\dag\sigma^jD^\mu D^\rho\Phi\big)( D_\mu W_{\nu\rho})^j+{\rm H.c.},
\end{equation}
with the coefficients $\overline{c}_{D\Phi}$ and $\overline{c}_{DW}$ dimensionless, so that both Lagrangian terms ${\cal L}_{D\Phi}$ and ${\cal L}_{DW}$ are associated to mass-dimension-8 interactions. The occurrence of the spontaneous breaking of the electroweak gauge group yields
\begin{eqnarray}
&&
{\cal L}_{D\Phi}=\frac{\overline{c}_{D\Phi}}{m_W^3}W^{+\nu}\hspace{0.000001cm}_\rho W^-_{\nu\mu}\,\partial^\mu\partial^\rho h+\cdots,
\end{eqnarray}
\begin{eqnarray}
&&
{\cal L}_{DW}=
-\frac{\overline{c}_{DW}}{2m_W^3}
\Big(
W^{+\nu}\partial_\mu W^-_{\nu\rho}+{\rm H.c.}
\Big)\partial^\mu\partial^\rho h
\nonumber \\ && \hspace{0.4cm}
-\frac{\overline{c}_{DW}}{2m_W^3}
\Big(
\partial^\mu W^{+\rho}\,\partial_\mu W^-_{\nu\rho}+{\rm H.c.}
\Big)\partial^\nu h+\cdots,
\label{LDWasb}
\end{eqnarray}
with the ellipses indicating the presence of further terms, corresponding to couplings other than $WWh$. While a proper discussion on $WWh$ couplings emerged from mass-dimension-8 Lagrangian terms should include the determination of an operator basis, the goal of Eqs.~\eqref{LPhibsb}-\eqref{LDWasb} is rather to show how $\lambda_4$ and $\lambda_5$ couplings, in Eq.~\eqref{LCPeven}, can be generated by effective-Lagrangian terms with mass-dimensions $\geqslant8$. On the other hand, the Lagrangian ${\cal L}^{\rm odd}_{WWh}$, which violates $CP$ symmetry, includes two terms, of which the one involving $\tilde\lambda_1$ is generated by a $({\rm mass})^6$-dimension SM effective Lagrangian term comprehended by ${\cal L}_{\rm SILH}$, as discussed above. As far as the $\tilde\lambda_2$ term is concerned, the number of derivatives involved in its structure suggests that can be generated by an ${\rm SU}(2)_L\otimes{\rm U}(1)_Y$-invariant effective-Lagrangian term of mass dimension 6. While the SILH basis, used for the present discussion, does not come with such a term, thus suggesting that the $\tilde{\lambda}_2$ coupling might emerge from a term of dimension $\big( {\rm mass} \big)^8$ or larger, we consider, just for illustrative purposes, the mass-dimension-6 
\begin{eqnarray}
&&
\tilde{\cal L}_{DW\Phi}=\frac{\tilde{c}_{DW\Phi}}{2m_W^2}
\big(( D^\nu\Phi )^\dag \sigma^j\Phi \big)\big( D^\mu W_{\mu\nu} \big)^j+{\rm H.c.}
\nonumber \\ && \hspace{0.7cm}
=\frac{\tilde{c}_{DW\Phi}}{m_W}
\Big(
i\,W^{+\nu}\partial^\mu W^-_{\mu\nu}+{\rm H.c.}
\Big) h
\nonumber \\ && \hspace{0.7cm}
+\frac{\tilde{c}_{DW\Phi}}{m_W}
\Big(
iW^-_\nu\partial^\mu W^+_\mu+{\rm H.c.}
\Big)\partial^\nu h
\nonumber \\ && \hspace{0.7cm}
+\frac{\tilde{c}_{DW\Phi}}{m_W}
\Big(
i\,W^{-\mu}\partial_\nu W^+_\mu+{\rm H.c.}
\Big)\partial^\nu h
\nonumber \\ && \hspace{0.7cm}
+\frac{2\,\tilde{c}_{DW\Phi}}{m_W}
\Big(
i\,W^+_\mu\partial^\mu W^-_\nu+{\rm H.c.}
\Big)\partial^\nu h+\cdots,
\label{offshellCPodd}
\end{eqnarray}
which, after electroweak symmetry breaking, comes along with the $CP$-violating $\tilde{\lambda}_2$ coupling. Note, however, that the addition of a factor $\Phi^\dag\Phi$ to $\tilde{{\cal L}}_{DW\Phi}$ defines a mass-dimension-8 term which also generates $\tilde{\lambda}_2$.
\\

\begin{figure}[ht]
\center
\includegraphics[width=5cm]{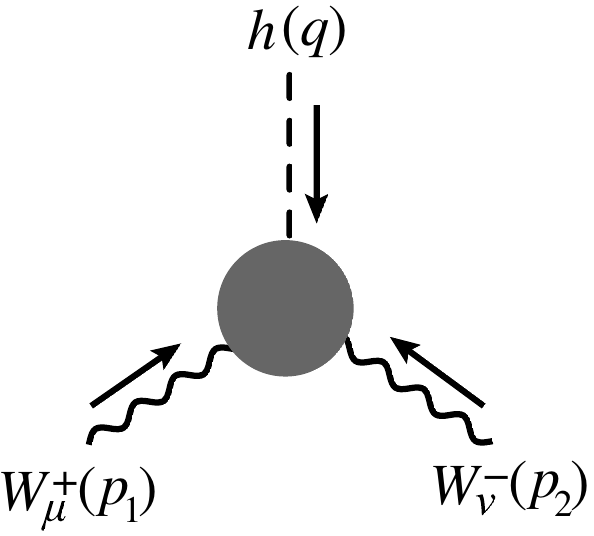}
\caption{\label{WWhconventions} Conventions for the momenta and for the Lorentz indices adopted for both the vertex function associated to ${\cal L}^{\rm eff}_{WWh}$ and the one-loop calculation of the vertex function $WWh$.}
\end{figure}

Now we use the conventions shown in Fig.~\ref{WWhconventions}
to write down the vertex function for ${\cal L}^{\rm eff}_{WWh}$. This vertex function, which we denote by $\Gamma^{\rm eff}_{\mu\nu}$, is derived under the assumption that the three external particles are off the mass shell. Since the momentum-conservation condition $p_1+p_2+q=0$ holds, we can write the vertex function in terms of two independent 4-momenta, which we chose to be $p_1$ and $p_2$. As we did with the effective Lagrangian ${\cal L}_{WWh}^{\rm eff}$, we split this vertex function as the sum of its $CP$-even and $CP$-odd contributions, namely, $\Gamma^{\rm eff}_{\mu\nu}=\Gamma^{\rm even}_{\mu\nu}+\Gamma^{\rm odd}_{\mu\nu}$, with the vertex-function terms given as 
\begin{eqnarray}
&&
\Gamma^{\rm even}_{\mu\nu}=g
\bigg(
m_W\lambda_1g_{\mu\nu}+\frac{\lambda_2}{m_W}\big( p_{1\nu}p_{2\mu}-g_{\mu\nu}\,p_1\cdot p_2 \big)
\nonumber \\ &&
\hspace{0.55cm}
-\frac{\lambda_3}{m_W}\big( p_{1\mu}p_{1\nu}+p_{2\mu}p_{2\nu}-g_{\mu\nu}p_1^2-g_{\mu\nu}p_2^2 \big) 
\nonumber \\ &&
\hspace{0.55cm}
-\frac{\lambda_4}{m_W^3}
\Big( p_{1\mu}p_{1\nu}\big( p_1^2+p_1\cdot p_2 \big)+p_{2\mu}p_{2\nu}\big( p_2^2+p_1\cdot p_2 \big)
\nonumber \\ &&
\hspace{0.55cm}
-g_{\mu\nu}\big( \big( p_2^2+p_1\cdot p_2 \big)^2+\big( p_1^2+p_1\cdot p_2 \big)^2 \big)
\nonumber \\ &&
\hspace{0.55cm}
+p_{1\nu}p_{2\mu}(p_1+p_2)^2  
\Big)
+\frac{\lambda_5}{m_W^3}\Big( p_2^2\,p_{1\mu}p_{1\nu}
\nonumber \\ &&
\hspace{0.55cm}
+p_1^2\,p_{2\mu}p_{2\nu}-(p_1\cdot p_2)p_{1\mu}p_{2\nu}-p_1^2\,p_2^2\,g_{\mu\nu}
\nonumber \\ &&
\hspace{0.55cm}
-(p_1^2+p_1\cdot p_2+p_2^2)(p_1\cdot p_2g_{\mu\nu}-p_{1\nu}p_{2\mu}) \Big)
\bigg),
\label{Geven}
\end{eqnarray}
\begin{eqnarray}
&&
\Gamma_{\mu\nu}^{\rm odd}=g
\Big(
\frac{\tilde\lambda_1}{m_W}\epsilon_{\mu\nu\alpha\beta}\,p_2^\alpha p_1^\beta
\nonumber \\ &&
\hspace{0.3cm}
+\frac{i\tilde\lambda_2}{m_W}
\big(
g_{\mu\nu}p_2^2-p_{2\mu}p_{2\nu}-g_{\mu\nu}p_1^2+p_{1\mu}p_{1\nu}
\big)
\Big).
\label{Godd}
\end{eqnarray}
It can be appreciated, from Eq.~(\ref{Godd}), that the $CP$-violating vertex function $\Gamma^{\rm odd}_{\mu\nu}$ is antisymmetric with respect to the interchange $p_1\longleftrightarrow p_2$. On the other hand, $\Gamma^{\rm even}_{\mu\nu}$, which is $CP$ preserving, does not have a definite symmetry property with respect to such an interchange, though some of its terms are symmetric with respect to the aforementioned momenta interchange. 
\\

\subsection{One-loop contributions to \texorpdfstring{$WWh$}\text{\hspace{2mm}from virtual neutrinos}}
We now address the contributions from Majorana neutrinos, in the framework of the seesaw variant given in Ref.~\cite{Pilaftsis:1991ug}, to the $WWh$ vertex at one loop. To write down the relevant expressions, from here on we follow the notation of Refs.~\cite{Martinez:2022epq,Novales-Sanchez:2023ztg,Novales-Sanchez:2024pso,Ramirez:2025zyg} and, in accordance, use the generic notation $n_j$ to refer to the whole set of neutrino fields, both light ones and heavy ones, with $n_1=\nu_1$, $n_2=\nu_2$, $n_3=\nu_3$, $n_4=N_1$, $n_5=N_2$, and $n_6=N_3$. Moreover, for the one-loop calculation of the $WWh$ vertex function we adopt the conventions given in Fig.~\ref{WWhconventions}. The one-loop contribution to the $WWh$ vertex can be written as the sum
\begin{equation}
\Gamma^{WWh}_{\mu\nu}=\Gamma_{\mu\nu}^{\ell\ell n}+\Gamma_{\mu\nu}^{\ell nn}+\Gamma_{\mu\nu}^{nn},
\end{equation}
with the contributing terms $\Gamma_{\mu\nu}^{\ell\ell n}$, $\Gamma_{\mu\nu}^{\ell nn}$, and $\Gamma_{\mu\nu}^{nn}$ diagrammatically expressed as
\begin{equation}
\Gamma_{\mu\nu}^{\ell\ell n}=
\sum_\alpha\sum_{j=1}^6
\begin{gathered}
\vspace{-0.2cm}
\includegraphics[width=1.7cm]{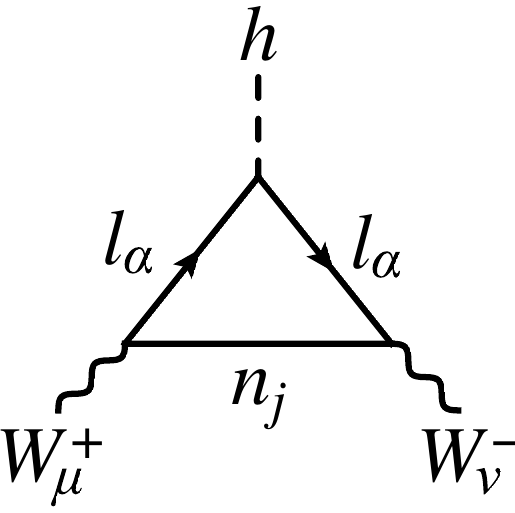}
\end{gathered}
,
\label{Glln}
\end{equation}
\begin{equation}
\Gamma_{\mu\nu}^{\ell nn}=
\sum_\alpha\sum_{j=1}^6\sum_{k=1}^6
\Bigg(
\hspace{0.2cm}
\begin{gathered}
\vspace{-0.2cm}
\includegraphics[width=1.7cm]{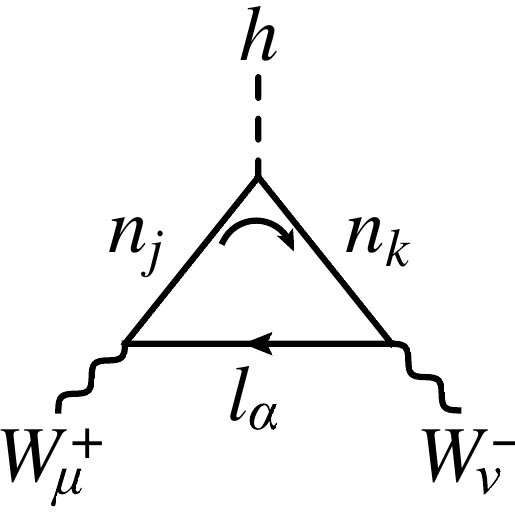}
\end{gathered}
+
\begin{gathered}
\vspace{-0.2cm}
\includegraphics[width=1.7cm]{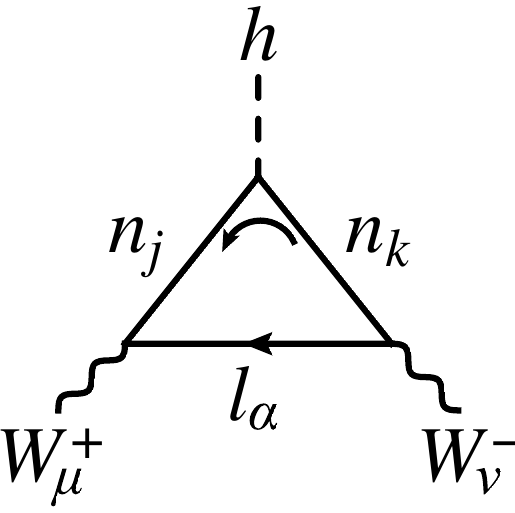}
\end{gathered}
\hspace{0.2cm}
\Bigg),
\label{Glnn}
\end{equation}
\begin{equation}
\Gamma_{\mu\nu}^{nn}=
\sum_{j=1}^6\sum_{k=1}^6
\Bigg(
\hspace{0.2cm}
\begin{gathered}
\vspace{-0.2cm}
\includegraphics[width=1.7cm]{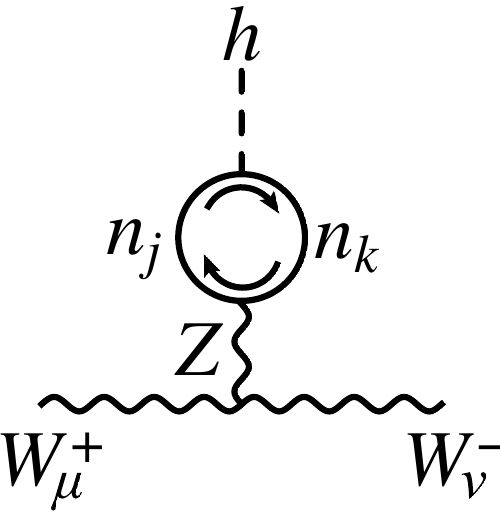}
\end{gathered}
+
\begin{gathered}
\vspace{-0.2cm}
\includegraphics[width=1.7cm]{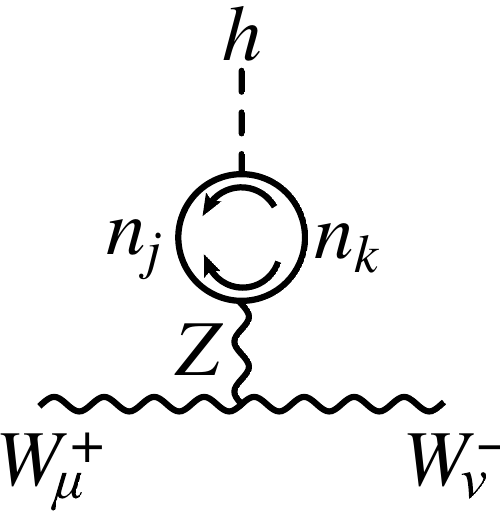}
\end{gathered}
\hspace{0.2cm}
\Bigg).
\label{Gnn}
\end{equation}
Eqs.~(\ref{Glln})-(\ref{Gnn}), illustrate how the set of contributing Feynman diagrams follows from neutrino couplings $W\ell_\alpha n_j$, $n_jn_kh$, and $n_jn_kZ$, which are associated to the new physics behind neutrino-mass generation, whereas all other couplings come from the SM. The involved new-physics neutrino couplings can be checked out in Eqs.~(\ref{LWnl})-(\ref{LNC}). Differences between Dirac and Majorana neutrinos manifest at different levels, including calculational aspects, as it is the case of the Feynman rules~\cite{Feynman:1949zx,Gluza:1991wj,Denner:1992vza}. The Majorana framework, as compared to the Dirac case, usually brings a larger number of contributing diagrams. From the viewpoint of Wick's theorem~\cite{Wick:1950ee}, this is an outcome of the larger allowed number of contractions involving Majorana fields. In this regard, two different vertices for couplings $n_jn_kh$ are to be taken into account, and the same goes for $n_jn_kZ$ couplings. For instance, ${\cal L}_{h\nu\nu}$, which we display in Eq.~(\ref{Lhnn}), has the form ${\cal L}_{h\nu\nu}=-i\sum_k\sum_jh\,\overline{n_k}\,\Gamma_{kj}n_j$, from which the Feynman rules
\begin{eqnarray}
\begin{gathered}
\vspace{-0.2cm}
\includegraphics[width=1.7cm]{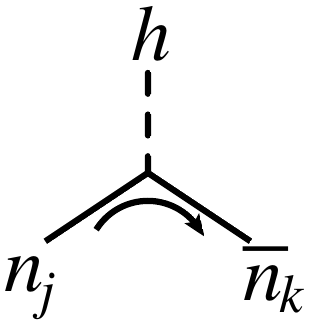}
\end{gathered}
&=&\Gamma_{kj},
\label{FeynruleD}
\\ \nonumber \\ \nonumber \\
\begin{gathered}
\vspace{-0.2cm}
\includegraphics[width=1.7cm]{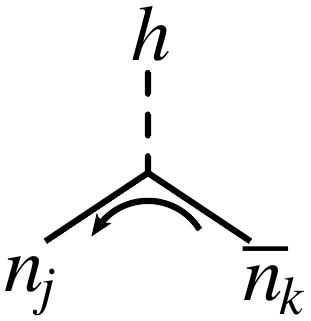}
\end{gathered}
&=&\Gamma'_{kj}=C\,\Gamma_{jk}^{\rm T}C^{-1},
\label{FeynruleM}
\end{eqnarray}
are obtained, where $\Gamma^{\rm T}_{jk}$ is the matrix transpose of $\Gamma_{kj}$. The Feynman rule shown in Eq.~(\ref{FeynruleD}) would be present in both the Dirac and the Majorana cases, whereas the Feynman rule of Eq.~(\ref{FeynruleM}) is exclusive to Majorana neutrinos. The difference in the diagrammatic representations of such rules lies in the arrows off of fermion lines, which represent a reference fermion flow~\cite{Denner:1992vza}. As shown in Eq.~(\ref{Glnn}), the $\Gamma_{\mu\nu}^{\ell nn}$ contribution involves the sum of two generic diagrams, which are then summed over neutrino types, labeled by $k,j=1,2,3,4,5,6$, and lepton flavors, $\alpha=e,\mu,\tau$. Observe that one of these diagrams involves a $\overline{n_k}n_jh$ vertex given by Eq.~\eqref{FeynruleD}, whereas the other includes a $\overline{n_k}n_jh$ vertex corresponding to Eq.~\eqref{FeynruleM} instead. The structure of $\Gamma^{\rm T}_{ij}$ and the properties of the charge-conjugation matrix $C$ conspire to yield
\begin{equation}
\Gamma'_{kj}=\Gamma_{kj},
\end{equation}
that is, the Feynman rules given in Eq.~(\ref{FeynruleD}) and (\ref{FeynruleM}) are indeed the same, thus implying that the two diagrams in Eq.~(\ref{Glnn}) are equal to each other. By the same token, the contributing diagrams in Eq.~(\ref{Gnn}) are the same. A similar discussion applies to the vertex $\overline{n_k}n_jZ$, of which, again, there are two versions. 
\\

UV divergences are a latent issue when dealing with loop integrals, which thus calls for a regularization scheme, among which we choose dimensional regularization~\cite{Bollini:1972ui,tHooft:1972tcz}. Usage of dimensional regularization has the advantages of getting well along with gauge symmetry, as well as the standardization of its implementation in software tools. In this approach, 4-momentum integrals are tackled under the assumption that the dimension of spacetime is $D$, where 1 dimension is time-like and the remaining $D-1$ dimensions are space-like. The resolution of loop integrals is carried out in some spacetime with $D$ dimensions in which such integrals are convergent. Together with this assumption on the number of spacetime dimensions, 4-momentum integrals change from $\int d^4k\,f(k)$ to $\mu_{\rm R}^{4-D}\int d^Dk\,f(k)$, for some function $f(k)$, of loop momentum, and where $\mu_{\rm R}$, known as the renormalization scale, is a quantity with units of mass, which corrects the units of these integrals in order for them to preserve their mass dimensions for whichever $D$ is considered. At some point, an analytic continuation is implemented on the loop integrals by assuming that the spacetime dimension $D$ is a complex number, then defining the quantity $\epsilon=4-D$, with $\epsilon\to0$. Along the process, divergent integrals are written as a sum of a term proportional to $\Delta_\epsilon=\frac{2}{\epsilon}-\gamma_{\rm EM}+\log\,4\pi$ ($\gamma_{\rm EM}$ is the Euler-Mascheroni constant), which thus formally diverges as $\epsilon\to0$, and a finite contribution. A useful way to determine whether loop integrals might be divergent is the superficial degree of divergence (SDD), defined as the difference of the maximum power of loop momentum in the numerator minus the maximum power of loop momentum in the denominator. Whenever ${\rm SDD}<0$, the momentum integral is finite for sure, but if, in contraposition, ${\rm SDD}\geqslant0$, then UV divergences may arise. Notice, however, that a non-negative SDD does not imply the presence of divergencies, since symmetries could render the expressions finite, upon addition of the whole set of contributions in a given calculation. The SDD of the diagrams of Eq.~(\ref{Gnn}) is 2, whereas the SDD for the diagrams contributing to $\Gamma^{\ell\ell n}$ and $\Gamma^{\ell nn}$, Eqs.~(\ref{Glln}) and (\ref{Glnn}), is 1, therefore indicating that UV divergences originating in all the contributing diagrams are expected. Moreover, the existence of the vertex $WWh$ in the SM at the tree level is another well-motivated reason for UV-divergent contributions to manifest, concretely in the contributions to the coefficient $\lambda_1$, in Eq.~(\ref{Geven}). This is in fact the case. Contributions to all other factors $\lambda_j$ and $\tilde\lambda_j$, in Eqs.~(\ref{Geven}) and (\ref{Godd}), are, by the same criterion, expected to be free of UV divergences. Since we are interested in the $WWh$ AC, which are not present in the SM at the tree level, we disregard the new-physics loop contributions to $\lambda_1$. In what follows, we respectively denote by $\lambda_j^{\rm NP}$ and $\tilde\lambda_j^{\rm NP}$ the new-physics contributions to the $\lambda_j$ and $\tilde\lambda_j$ factors displayed in Eqs.~(\ref{Geven}) and (\ref{Godd}).
\\

A widely standardized mean, in view of its remarkable practicality, to address loop integrals is the tensor-reduction method, by Passarino and Veltman~\cite{Passarino:1978jh}. In this method, the general form of the solutions to tensor momentum integrals are sort of guessed by decomposing them into combinations of Lorentz tensor structures, which then leads to expressions for the loop integrals in terms of scalar momentum integrals characterized by the general form,
\begin{equation}
T^{(N)}_0({\bf p},{\bf m})=\frac{(2\pi\mu_{\rm R})^{4-D}}{i\pi^2}\int d^Dk\prod_{A=0}^{N-1}\big( (q+p_A)^2-m_A^2 \big)^{-1},
\end{equation}
with $p_0=0$ and where we have denoted ${\bf p}=\big( p_1,p_2,\ldots,p_{N-1} \big)$ and ${\bf m}=\big( m_0,m_1,\ldots,m_{N-1} \big)$. $T^{(N)}_0\big( {\bf p},{\bf m} \big)$ is dubbed the $N$-point Passarino-Veltman scalar function, or $N$-point scalar function, for short. The 1-point and 2-point scalar functions, respectively denoted by $A_0$ and $B_0$, are UV divergent, whereas the 3-point scalar function, referred to by $C_0$, as well as all other scalar functions with $N>2$ are UV finite. To carry out the calculation of the contributions from the Feynman diagrams of Eqs.~(\ref{Glln})-(\ref{Gnn}), we have used the software implementations of the tensor-reduction method in the packages \textsc{FeynCalc}~\cite{Mertig:1990an,Shtabovenko:2016sxi,Shtabovenko:2020gxv} and \textsc{Package-X}~\cite{Patel:2015tea}, for \textsc{Mathematica} by Wolfram. 
\\

Consider the generic diagram contributing to $\Gamma_{\mu\nu}^{nn}$, which has been given in Eq.~(\ref{Gnn}). The calculation of the analytic expression for this diagram results in
\begin{equation}
\begin{gathered}
\vspace{-0.2cm}
\includegraphics[width=1.7cm]{v1/WWhdiag31.pdf}
\end{gathered}
=\big( g^{\mu\nu}p_2^2-p_{2\mu}p_{2\nu}-g_{\mu\nu}p_1^2+p_{1\mu}p_{1\nu} \big) F_{kj},
\end{equation}
where $F_{kj}$ is some function of masses and momenta, expressed in terms of Passarino-Veltman scalar functions. Notice that the Lorentz tensor structure characterizing this contributions is exactly the one associated to $\tilde\lambda_2$, shown in Eq.~(\ref{Godd}) and which is $CP$ nonpreserving. However, we have reached the conclusion that, even though the calculation has been preformed with the $WWh$ vertex off shell, the function $F_{kj}$ vanishes exactly. Therefore, the total contribution $\Gamma_{\mu\nu}^{nn}$, displayed in Eq.~(\ref{Gnn}), is zero. 
\\

The remaining contributions from the virtual Majorana neutrinos are those of Eqs.~(\ref{Glln}) and (\ref{Glnn}). These generate a total contribution matching the Lorentz-covariant structure of the effective vertex functions $\Gamma^{\rm even}_{\mu\nu}$ and $\Gamma^{\rm odd}_{\mu\nu}$, explicitly displayed in Eqs.~(\ref{Geven})-(\ref{Godd}). Our calculation shows that 1-loop contributions to $\lambda_2$ are absent, that is, $\lambda_2^{\rm NP}=0$, whereas all other anomalous form-factor contributions $\lambda_j^{\rm NP}$ and $\tilde\lambda_j^{\rm NP}$ are generated. Moreover, we find that $\lambda_4^{\rm NP}=\lambda_5^{\rm NP}$ holds. We jointly denote the $CP$-even and $CP$-odd contributions to the $WWh$ couplings as $L^{\rm NP}=\lambda_j^{\rm NP},\tilde\lambda_j^{\rm NP}$. It turns out that any of these form factors can be accommodated as
\begin{eqnarray}
&&
L^{\rm NP}=\sum_{k=1}^6\sum_\alpha|{\cal B}_{\alpha n_k}|^2L^{(1)}_{\alpha n_k}
\nonumber \\ &&
\hspace{0.9cm}
+\sum_{j=1}^6\sum_{k=1}^6\sum_\alpha{\cal B}_{\alpha n_k}{\cal B}_{\alpha n_j}^*
\nonumber \\ &&
\hspace{0.8cm}
\times
\Big( 
{\cal C}_{n_kn_j}\big(
m_{n_j}L_{\alpha n_kn_j}^{(2)}+m_{n_k}L_{\alpha n_kn_j}^{(3)}
\big)
\nonumber \\ &&
\hspace{0.9cm}
+{\cal C}_{n_kn_j}^*
\big(
m_{n_k}L_{\alpha n_kn_j}^{(2)}+m_{n_j}L_{\alpha n_kn_j}^{(3)}
\big)
\Big),
\label{LNPhysics}
\end{eqnarray}
where $L^{(1)}_{\alpha n_k}$ and $L^{(a)}_{\alpha n_kn_j}$, with $a=2,3$, are functions of masses and off-shell external momenta, nested within 2-point and 3-point scalar functions. The term with only two sums, in the first line of the equation, represents the contribution from the vertex function $\Gamma^{\ell\ell n}_{\mu\nu}$, Eq.~(\ref{Glln}), whereas the other term, the one with three sums, comprehends the whole contribution from $\Gamma_{\mu\nu}^{\ell nn}$, Eq.~(\ref{Glnn}). The contributions we have found for the $L^{(1)}_{\alpha n_k}$ and $L^{(a)}_{\alpha n_k n_j}$ factors can be written, in general, as
\begin{equation}
L^{(a)}=\sum_n\beta^{(a,n)} \Delta B_0^{(a,n)}+\sum_m\gamma^{(a,m)}C_0^{(a,m)}.
\end{equation}
Here, $C_0^{(a,m)}$ stands for a 3-point scalar function, whereas $\Delta B_0^{(a,n)}$ is, for each $n$, a difference $B_0(p_k,m_{k_1},m_{k_2})-B_0(p_j,m_{j_1},m_{j_2})$, of two $B_0$ functions. The sums in this equation indicate the presence of 2-point and 3-point functions, each of them accompanied by a coefficient $\beta^{(a,n)}$ or $\gamma^{(a,m)}$, which are simple functions of masses and scalar products of momenta $p_1$ and $p_2$. Any 2-point function can be split as a sum $B_0(p_k,m_{k_1},m_{k_2})=\Delta_\epsilon+\log\,\mu_{\rm R}^2+f(p_k,m_{k_1},m_{k_2})$, where $\Delta_\epsilon$ is the UV-divergent term, previously introduced, and $f(p_k,m_{k_1},m_{k_2})$ is free of UV divergences. As our notation indicates, all the $B_0$ functions, regardless of their specific arguments, share the very same divergent term $\Delta_\epsilon$, and the $\mu_{\rm R}$ is also the same. Therefore, both UV divergences and the $\mu_{\rm R}$ dependence cancel out from differences $\Delta B_0^{(a,n)}$ and, since $C_0$ functions are finite, we conclude that all the NP contributions $L^{(a)}$ are free of UV divergences. Further, these contributions are also independent of the renormalization scale $\mu_{\rm R}$.
\\

Regarding Eq.~\eqref{LNPhysics}, an aspect worth of discussion, for the sake of comparison, is what the analogous AC contributions produced in the framework of the original type-1 seesaw (T1SS) mechanism would be. Just for a moment, let us consider the corresponding T1SS contribution, $L^{\rm T1SS}$, to any AC, for which we first point out that the structure of $L^{\rm T1SS}$ is almost the same as that of Eq.~\eqref{LNPhysics}. The only difference is that, while in the case of the model of radiatively-induced light-neutrino masses of Ref.~\cite{Pilaftsis:1991ug} the neutrino-mass factors $m_{n_j}$ in the last two lines of this equation vanish if $n_j=\nu_j$, in the T1SS, by contrast, such mass factors must be kept. The matrix elements ${\cal B}_{\alpha n_k}$ and $C_{n_kn_j}$, appearing in Eq.~\eqref{LNPhysics}, and thus in the corresponding T1SS contribution, can be expressed in terms of a $3\times3$ matrix $\xi$, which, under the assumption that the new-physics scale $\Lambda$ is large, can be approximated as $\xi\simeq \frac{v}{\Lambda}Y_{\rm MD}$. Here, $Y_{\rm MD}$ is some $3\times3$ matrix, which we reasonably assume to be ${\cal O}(1)$. Furthermore, we express T1SS neutrino masses as $m_{\nu_j}=\frac{v^2}{\Lambda}Y_{\nu_j}$ and $m_{N_j}\simeq\Lambda Y_{N_j}$, where both $Y_{\nu_j}$ and $Y_{N_j}$ are also considered to be ${\cal O}(1)$. With these elements at hand, we write the $L^{\rm T1SS}$ contribution as
\begin{equation}
L^{\rm T1SS}\simeq\sum_\alpha\sum_{k=1}^3
\Big(
F^{(0)}_{\alpha k}+\frac{v^2}{\Lambda}F^{(1)}_{\alpha k}+\frac{v^2}{\Lambda^2}F^{(2)}_{\alpha k}
\Big),
\label{LT1SS}
\end{equation}
with the following definitions:
\begin{equation}
F^{(0)}_{\alpha k}=|\big( U_{\rm PMNS} \big)_{\alpha k}|^2L^{(1)}_{\alpha \nu_k},
\label{F0}
\end{equation}
\begin{eqnarray}
&&
F^{(1)}_{\alpha k}=2|\big( U_{\rm PMNS} \big)_{\alpha k}|^2Y_{\nu_k}
\Big(
L^{(2)}_{\alpha \nu_k\nu_k}+L^{(3)}_{\alpha \nu_k\nu_k}
\Big)
\nonumber \\ && \hspace{0.3cm}
+\big( U_{\rm PMNS} \big)_{\alpha k}\sum_{j=1}^3Y_{N_j}Z_{\alpha j}^* \Big( (Y_{\rm MD})_{kj}L^{(2)}_{\alpha\nu_k N_j}
\nonumber \\ && \hspace{1.2cm}
+(Y_{\rm MD})_{kj}^*L^{(3)}_{\alpha\nu_k N_j} \Big)
\nonumber \\ && \hspace{0.3cm}
+\big( U_{\rm PMNS} \big)_{\alpha k}^*\sum_{j=1}^3Y_{N_j}Z_{\alpha j}\big( (Y_{\rm MD})_{kj}L^{(2)}_{\alpha N_j\nu_k}
\nonumber \\ && \hspace{1.2cm}
+(Y_{\rm MD})_{kj}^*L^{(3)}_{\alpha N_j \nu_k} \big),
\nonumber \\
\label{F1}
\end{eqnarray}
\begin{eqnarray}
&&
F^{(3)}_{\alpha k}=-2{\rm Re}\{ \big( U_{\rm PMNS} \big)_{\alpha k}J_{\alpha k}^* \}
L^{(1)}_{\alpha \nu_k}
\nonumber \\ && \hspace{0.8cm}
+|Z_{\alpha k}|^2L^{(1)}_{\alpha N_k},
\label{F2}
\end{eqnarray}
with $U_{\rm PMNS}$ the Pontecorvo-Maki-Nakagawa-Sakata (PMNS) neutrino-mixing matrix. In order to write down Eqs.~\eqref{LT1SS}-\eqref{F2}, the matrices $Z=U_{\rm PMNS}Y_{\rm MD}$ and $J=U_{\rm PMNS}Y_{\rm MD}Y_{\rm MD}^\dag$ have been defined. About Eq.~\eqref{LT1SS}, note that the factors $F^{(a)}_{\alpha k}$ are $\Lambda$-scale dependent, which can be understood by taking into account that the factors $L^{(a)}$, comprised by them, depend on neutrino masses, and thus come along with $\Lambda$ dependence. For large $\Lambda$, the $L^{(a)}$ factors behave in such a way that
\begin{equation}
\lim_{\Lambda\to\infty}\frac{v^2}{\Lambda}F^{(1)}_{\alpha k}=0,
\label{limitF1}
\end{equation}
\begin{equation}
\lim_{\Lambda\to\infty}\frac{v^2}{\Lambda^2}F^{(2)}_{\alpha k}=0.
\label{limitF2}
\end{equation}
In the case of $F^{(0)}_{\alpha k}$, a nonzero value is reached in the limit as $\Lambda\to\infty$. Since $m_{\nu_k}\to0$ as $\Lambda\to\infty$ for all $k=1,2,3$, we have
\begin{equation}
\lim_{\Lambda\to\infty}L^{(1)}_{\alpha \nu_k}=L^{(1)}_\alpha,
\end{equation}
with $L^{(0)}_\alpha$ finite and independent of mass-eigenspinor indices $k=1,2,3$. Therefore,
\begin{equation}
\lim_{\Lambda\to\infty}F^{(0)}_{\alpha k}=\sum_\alpha L^{(1)}_\alpha\sum_{k=1}^3\big(U_{\rm PMNS}U_{\rm PMNS}^\dag\big)_{\alpha\alpha}
=\sum_\alpha L^{(1)}_\alpha,
\end{equation}
in which no remnant of light-neutrino mixing nor massiveness remains. As previously discussed, AC contributions come from the triangle diagrams displayed in Eqs.~\eqref{Glln} and \eqref{Glnn}. In particular, the factor $L^{(1)}_{\alpha\nu_k}$ originates in the diagrams of Eq.~\eqref{Glln}, in which no vertex $h\nu_j\nu_k$ is involved; in other words, these diagrams need no couplings of neutrinos with the Higgs field. This is consistent with the fact that the vanishing of light-neutrino masses eliminates the couplings of neutrinos to the Higgs boson. All these aspects lead us to the conclusion that the contribution $L^{\rm T1SS}$ decouples as $\Lambda\to\infty$~\cite{Appelquist:1974tg}, in which case the resultant contribution corresponds to the one from massless neutrinos. This decoupling behavior becomes quite relevant in the the presence of the enormous new-physics energy scale $\Lambda\sim10^{13}\,{\rm GeV}$, which characterizes the T1SS neutrino-mass mechanism, since effects from heavy neutrinos are expected to be dramatically suppressed and the contribution is rather determined by light neutrinos. From a quantitative viewpoint, this makes a noticeable difference with the contributions from the model of Ref.~\cite{Pilaftsis:1991ug}, in which the heavy neutrinos give rise to the leading effects, larger than those from light neutrinos by $\sim4$ orders of magnitude.



\section{Estimations and discussion}
\label{thenums}
For this section, our aim is a quantitative discussion of contributions from the new physics associated to the neutrino-mass generating mechanism of Ref.~\cite{Pilaftsis:1991ug} to the afore-discussed $WWh$ AC. Our analytical calculation, which has been described in the previous section, was carried out by taking the general case of three external lines off the mass shell. Across the present section, we consider particular cases of such a general result, determined by which of the external lines are assumed to be on shell. Each of these cases is contextualized in some physical process, which serves us to carry out our estimations of the contributions to the AC. 
\\ 

As displayed in Eq.~(\ref{LNPhysics}), the anomalous form-factor contributions, $\lambda_j^{\rm NP}$ and $\tilde\lambda_j^{\rm NP}$, depend on the matrices ${\cal B}$ and ${\cal C}$, Eqs.~(\ref{Bmatrix}) and (\ref{Cmatrix}), which, as shown in Ref.~\cite{Pilaftsis:1991ug}, are given in terms of the $3\times3$ complex matrix $\xi$ and the PMNS mixing matrix $U_{\rm PMNS}$. In order to reduce the number of parameters, with the objective of getting some estimation of effects, we write the matrix $\xi$ as $\xi=\rho X$, where $X$ is a $3\times3$ complex matrix, whose largest entry has modulus 1, and $\hat\rho$ is a real quantity characterizing the size of $\xi$. Then, we assume the matrix texture $X\approx e^{i\phi}\cdot{\bf 1}_3$, with ${\bf 1}_3$ the $3\times3$ identity matrix. So, we have
\begin{equation}
\xi\approx\hat\rho\,e^{i\phi}\cdot{\bf 1}_3.
\label{xitexture}
\end{equation}
While this texture is evidently not general, let us point out that other matrix textures do not dramatically change our quantitative results. Further, this form of $\xi$ avoids severe constraints on $\hat\rho$ coming from lepton-flavor-violation decays $l_\alpha\to l_\beta\gamma$~\cite{Ramirez:2025zyg}, which in turn allows us to get larger contributions to the AC $\lambda_j^{\rm NP}$ and $\tilde\lambda_j^{\rm NP}$. About $U_{\rm PMNS}$, this matrix can, in general, be written as $U_{\rm PMNS}=U_{\rm D}\,U_{\rm M}$, where $U_{\rm D}$ as well as $U_{\rm M}$ are $3\times3$ unitary matrices. If light neutrinos were described by Dirac fields, $U_{\rm M}$ would be absent, so $U_{\rm PMNS}=U_{\rm D}$, whereas, on the other hand, both $U_{\rm D}$ and $U_{\rm M}$ play a role when light neutrinos are Majorana particles. The matrix $U_{\rm D}$ is usually parametrized by 3 mixing angles, denoted by $\theta_{12}$, $\theta_{23}$, and $\theta_{13}$, and a complex phase $\delta_{\rm D}$, called ``Dirac phase''. In the case of the matrix $U_{\rm M}={\rm diag}\big( 1, e^{i\phi_{{\rm M},1}}, e^{i\phi_{{\rm M},2}} \big)$, the phases $\phi_{{\rm M},1}$ and $\phi_{{\rm M},2}$ are known as ``Majorana phases''. For the present work, we take, for the sake of simplicity, $\phi_{{\rm M},1}=0$, $\phi_{{\rm M},2}=0$. For the rest of the parameters constituting $U_{\rm PMNS}$, we use the values recommended by the Particle Data Group~\cite{ParticleDataGroup:2024cfk}, based on the results presented in Refs.~\cite{Super-Kamiokande:2023jbt,MINOS:2020llm,NOvA:2021nfi,T2K:2023smv,IceCubeCollaboration:2023wtb,Super-Kamiokande:2023ahc,KM3NeT:2024ecf,RENO:2018dro,RENO:2019otc,DoubleChooz:2019qbj,DayaBay:2024hrv,NOvA:2023iam}, by a number of experimental collaborations. Then, we have the following values for the mixing angles:
\begin{eqnarray}
&&
\sin^2\theta_{12}=0.307\pm0.013,
\\ \nonumber \\ &&
\sin^2\theta_{23}=0.546\pm0.0021,
\\ \nonumber \\ &&
\sin^2\theta_{13}=0.0220\pm0.0007.
\end{eqnarray}
And for the Dirac phase, we use 
\begin{equation}
\delta_{\rm D}=-\frac{\pi}{2}.
\end{equation}
\\

Neutrino oscillations have led to the conclusion that neutrinos are massive, and yet the absolute scale of neutrino mass remains so far undetermined. Cosmological observations have been used to set the upper limit $\sum_jm_{\nu_j}<0.12\,{\rm eV}$, at $95\%{\rm C.L.}$~\cite{eBOSS:2020yzd,Planck:2018vyg}. The neutrinoless double beta decay, an elusive process allowed exclusively if neutrinos are Majorana fermions, has never been observed, even though a number of experimental groups have been pursuing it for years. The null results have rather translated into upper bounds, lying within $10^{-2}\,{\rm eV}-10^{-1}\,{\rm eV}$, on the so-called ``neutrino effective mass'', $m_{\beta\beta}=\big| \sum_j (U_{\rm PMNS})_{ej}^2\,m_{\nu_j} \big|$, which has been achieved by exploration of various isotopes~\cite{CUORE:2019yfd,GERDA:2020xhi,KamLAND-Zen:2016pfg}. Another restriction on neutrino mass, with the advantage of being independent of cosmological assumptions or whether neutrinos are Majorana or Dirac, has been reported in Ref.~\cite{KATRIN:2024cdt}, by the KATRIN Collaboration. According to this work, neutrino mass is upper bounded as $m_{\nu_j}\lesssim0.45\,{\rm eV}$. Strictly speaking, the masses of the three light neutrinos must differ of each other, which is a requirement of neutrino oscillations~\cite{Pontecorvo:1957cp}. Quadratic differences among neutrino masses, $\Delta m_{kj}=m_{\nu_j}^2-m_{\nu_k}^2$, have been measured with remarkable precision, finding that $\Delta m_{21}^2\sim10^{-5}\,{\rm eV}^2$~\cite{Super-Kamiokande:2023jbt} and $|\Delta m_{32}^2|\sim10^{-3}\,{\rm eV}^2$~\cite{KM3NeT:2024ecf,DayaBay:2024hrv,Super-Kamiokande:2023ahc,IceCubeCollaboration:2023wtb,T2K:2023smv,NOvA:2021nfi,MINOS:2020llm,RENO:2018dro,DayaBay:2022orm}, with the sign of the latter yet to be figured out, perhaps by next-generation facilities. Our numerical estimations show that light-neutrino masses produce suppressed contributions to the AC, whereas, on the other hand, the dominant contributions are rather determined by heavy-neutrino mass. In this context, the small differences among light-neutrino masses yield negligible quantitative variations. So, in favor of practicality, this set of masses are taken as degenerate: $m_{\nu_j}=m_{\nu_k}$ for any $j,k=1,2,3$. Moreover, for our estimations, we take the KATRIN result as our reference for the value of light-neutrino masses.
\\

As far as the heavy neutrinos are concerned, just recall that, in accordance with the model considered for the present investigation, the corresponding set of masses is restricted to be quasi-degenerate~\cite{Pilaftsis:1991ug}, so $m_{N_j}=m_N$ for $j=1,2,3$ and where $m_N$ is some reference mass. In previous investigations~\cite{Martinez:2022epq,Novales-Sanchez:2023ztg,Novales-Sanchez:2024pso}, by some of the authors of the present work, the $\hat\rho$ parameter has been given the values $0.58$ and $0.65$ for estimations, since such choices produce contributions large enough to enter sensitivity regions of future experimental facilities. With this in mind, we use $\hat\rho=0.65$ for the present work. According to an study by the CMS Collaboration~\cite{CMS:2018iaf}, this value of the $\hat\rho$ parameter constrains heavy-neutrino masses to be $m_N\gtrsim700\,{\rm GeV}$.
\\

Studies by the ATLAS Collaboration have established intervals where Higgs AC, coming from mass-dimension-6 terms which are part of the SM effective Lagrangian, should lie~\cite{ATLAS:2020fcp,ATLAS:2023mqy,ATLAS:2024pov,ATLAS:2025oiy,ATLAS:2022fnp,ATLAS:2023gzn}. The results provided by such works are usually presented in the so-called ``Warsaw basis''~\cite{Grzadkowski:2010es}, whose relation with the SILH basis has been discussed, for instance, in Ref.~\cite{Falkowski:2001958}, where a sort of dictionary, relating the corresponding sets of $CP$-even effective-Lagrangian coefficients, is provided. In the context of the SM effective field theory, a global fit, which takes into account data from the Large Electron-Positron collider and from the Large Hadron Collider (LHC), has been presented by the authors of Ref.~\cite{Ellis:2018gqa}, where Wilson coefficients characterizing the effective-Lagrangian terms that comprise the $CP$-preserving part of the SILH Lagrangian are constrained. In particular, that work includes the restrictions $\overline{c}_{\Phi W}=\big( 0.2\pm2.8 \big)\times10^{-2}$ and $\overline{c}_W=\big(-1.9\pm4.8\big)\times10^{-2}$, at $95\%$ CL. We thus have the following bounds on $CP$-even coefficients of our vertex-function parametrization:
\begin{equation}
|\lambda_2^{\rm global\,fit}|\leqslant6.0\times10^{-2},
\end{equation}
\begin{equation}
|\lambda_3^{\rm global\,fit}|\leqslant7.3\times10^{-2}.
\label{l3globalfit}
\end{equation}
The arrival of next-generation colliders is expected to improve sensitivity to Higgs-boson couplings. In Ref.~\cite{Denizli:2017pyu}, the investigation of Higgs production through $e^+e^-\to\nu\overline{\nu}h$, in the context of the CLIC working at a CME of $360\,{\rm GeV}$ led the authors of this work to conclude that such a machine will be able to establish constraints as stringent as
\begin{equation}
|\lambda_2^{\rm CLIC}|\leqslant1.3\times10^{-2},
\end{equation}
\begin{equation}
|\lambda_3^{\rm CLIC}|\leqslant1.2\times10^{-2}.
\label{l3CLIC}
\end{equation}
An estimation of the sensitivity of the HL-LHC to SILH coefficients has been provided in Ref.~\cite{Englert:2015hrx}, according to which upper bounds as restrictive as 
\begin{equation}
|\lambda_2^{\rm HL\textrm{-}LHC}|\leqslant8\times10^{-3},
\end{equation}
\begin{equation}
|\lambda_3^{\rm HL\textrm{-}LHC}|\leqslant8\times10^{-3},
\label{l3HLLHC}
\end{equation}
might be reached. Keep in mind that our loop calculation has yielded $\lambda_2^{\rm NP}=0$. As far as $CP$-violation effects are concerned, let us comment on Ref.~\cite{ATLAS:2019jst}, by the ATLAS Collaboration, where the cross section for the process $pp\to h\to\gamma\gamma$ is measured and analyzed, then finding agreement with SM predictions. In that work, advantage is taken from the achieved measurements to look for new-physics effects, using the SILH Lagrangian. One of the results reported in this reference is $\tilde{c}_{\Phi W}=\big( -0.1\pm6.4 \big)\times10^{-2}$, at $95\%$ CL, so we have 
\begin{equation}
|\tilde{\lambda}_1^{\rm ATLAS}|\leqslant1.3\times10^{-1}.
\label{l1ATLAS}
\end{equation}
In Ref.~\cite{Karadeniz:2019upm}, the expected sensitivity to $CP$-odd effects of the CLIC working at a CME of $3\,{\rm TeV}$ has been estimated, then claiming that this machine will be able to establish a bound as restrictive as
\begin{equation}
|\tilde{\lambda}_1^{\rm CLIC}|\leqslant1.4\times10^{-2}.
\label{l1CLIC}
\end{equation}
\\

We would like to get an estimation of the AC contributions, $\lambda_j^{\rm NP}$ and $\tilde{\lambda}_j^{\rm NP}$. These quantities depend on several parameters: masses, particularly the heavy-neutrino mass $m_N$; squared 4-momenta associated to $WWh$ external lines; the $3\times3$ matrix $\xi\approx\hat\rho\,e^{i\phi}\cdot{\bf 1}_3$, which essentially defines the matrices ${\cal B}$ and ${\cal C}$, introduced in Eqs.~\eqref{Bmatrix} and \eqref{Cmatrix}; and the PMNS matrix. Except for the squared external 4-momenta, we have already established how all these parameters are going to be treated. In order to have some reference about what values to consider for the squared 4-momenta, we have explored, for our analyses, two scenarios: (1) $W^*Wh$, that is, a $W$ boson is taken off mass shell, which we denote by $W^*$, whereas the other $W$ boson and the Higgs boson are assumed to be on shell; (2) $W^*W^*h$, where the Higgs boson is on shell, but the two $W$ bosons are off shell. These scenarios are discussed next, using physical processes for a proper contextualization.
\\


\subsection{Scenario 1: \texorpdfstring{$W^{*}Wh$}{W*Wh}}
\begin{figure}[ht]
\center
\includegraphics[width=4.5cm]{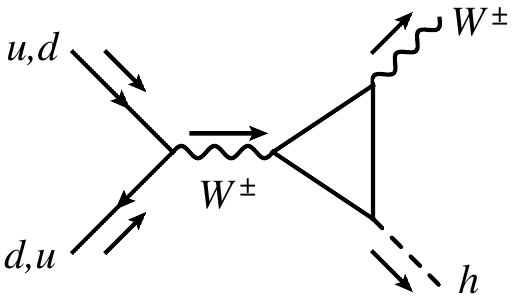}
\caption{\label{wwhcase1} Generic Feynman diagram illustrating how light and heavy neutrinos contribute to the subprocess $ud\to Wh$, at one loop, occurring $pp$ collisions.}
\end{figure}

We think of this situation in the framework of the process $pp\to Wh$, which receives contributions from subprocesses $ud\to Wh$ through $s$-channel diagrams, as shown in Fig.~\ref{wwhcase1}. 
The $Wh$ Higgs-production channel has great significance, since, together with $Zh$ production, it is the most effective mean for $h\to b\overline{b}$ decays detection. A thorough one-loop order calculation of $p\overline{p}\to Wh/Zh+X$, in the context of the SM, has been presented in Ref.~\cite{Ciccolini:2003jy}. Further, Refs.~\cite{ATLAS:2020fcp,ATLAS:2024yzu}, by the ATLAS Collaboration, reported measurements of $h\to b\overline{b}$ and even of $h\to c\overline{c}$ from $Wh$ and $Zh$ production. In what follows, we must distinguish two different CMEs: (1) the one corresponding to the $pp$ collision; and (2) the energy associated to the initial-state $u$ and $d$ quarks, which, being part of the colliding protons, take an energy fraction and give rise to the subprocess $ud\to Wh$. The latter CME is denoted, from now on, as $\sqrt{s}$. For our estimations and subsequent discussion on the AC contributions $\lambda_j^{\rm NP}$ and $\tilde{\lambda}_j^{\rm NP}$, we consider CMEs $\sqrt{s}$ ranging within $m_h+m_W\leqslant\sqrt{s}\leqslant39\,{\rm TeV}$. The lowest energy, $m_h+m_W$, is no other than the threshold for $Wh$ production. For the determination of the largest $\sqrt{s}$ value, we have used the \textsc{MadGraph} package~\cite{Alwall:2014hca} to simulate the process $pp\to Wh$, in order to have an estimation of the largest energy carried by the initial-state quarks involved in this process. For a $pp$ CME of $13\,{\rm TeV}$, corresponding to current LHC energies, we have $\sqrt{s}\approx5\,{\rm TeV}$. However, we have also considered, for our estimations, a future hadron collider, which, as claimed in Refs.~\cite{FCC:2018vvp,Bernardi:2022hny,Tang:2015qga,Tang:2022qku}, is aimed to reach $pp$ CMEs as large as $100\,{\rm TeV}$. In such case, simulations yield $\sqrt{s}\approx39\,{\rm TeV}$.
\\

\begin{figure}
\center
\includegraphics[width=7.5cm]{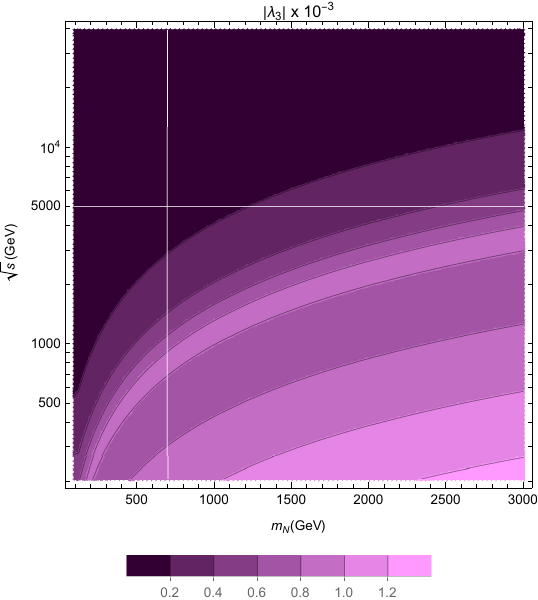}
\vspace{0.3cm}
\\
\includegraphics[width=7.5cm]{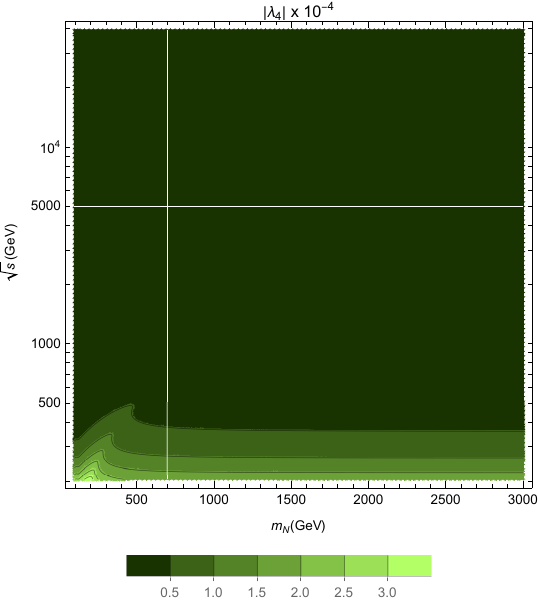}
\caption{\label{l34Sc1} Contributions, in the context of $Wh$ production, to $CP$-even AC $|\lambda_3^{\rm NP}|$ and $|\lambda_4^{\rm NP}|$, plotted in the $\big( m_N,\sqrt{s} \big)$ plane, for fixed $\hat{\rho}=0.65$. Values ranging within $100\,{\rm GeV}\leqslant m_N\leqslant3\,{\rm TeV}$ and $m_h+m_W\leqslant\sqrt{s}\leqslant39\,{\rm TeV}$ have been used. Contributions as large as $|\lambda_3^{\rm NP}|\sim10^{-3}$ and $|\lambda_4^{\rm NP}|\sim10^{-4}$ are generated.}
\end{figure}

Our quantitative discussion starts with the presentation of our estimations of $CP$-even anomalous $WWh$-coupling contributions, among which only $\lambda_3^{\rm NP}$ and $\lambda_4^{\rm NP}$ remain. While both AC are complex valued, we have rather estimated their moduli, that is, $|\lambda_3^{\rm NP}|$ and $|\lambda_4^{\rm NP}|$. We refer the reader to Fig.~\ref{l34Sc1},
where graphs depicting the corresponding contributions are displayed. The two graphs of Fig.~\ref{l34Sc1} have been plotted in the $\big( m_N,\sqrt{s} \big)$ parameter space, with $100\,{\rm GeV}\leqslant m_N\leqslant3\,{\rm TeV}$ and $m_h+m_W\leqslant\sqrt{s}\leqslant39\,{\rm TeV}$. The units used for these two parameters have been ${\rm GeVs}$. Both graphs display regions, colored in different tones, where the lighter the tone the larger the contribution. A labeling bar, placed beneath each graph, provides the size of the contributions corresponding to these regions, which have been scaled to orders $10^{-3}$ and $10^{-4}$ for $|\lambda_3^{\rm NP}|$ and $|\lambda_4^{\rm NP}|$, respectively. For the execution of the graphs, the $\sqrt{s}$ axes have been set in base-10 logarithmic scale. A horizontal line, at $\sqrt{s}\approx 5\,{\rm TeV}$, has been included, in each graph, to indicate the maximum $\sqrt{s}$, of the initial-state quarks, for $pp$ collisions occurring at a CME of $13\,{\rm TeV}$. A vertical line in each graph, at $m_N=700\,{\rm GeV}$, has been added, aiming at representing the smallest heavy-neutrino mass allowed by our choice $\hat\rho=0.65$, in accordance with the CMS paper on heavy-neutral-lepton masses~\cite{CMS:2018iaf}. It can be appreciated, from the graph in the upper panel of Fig.~\ref{l34Sc1}, that the values of the AC contribution $|\lambda_3^{\rm NP}|$ range from $\sim10^{-4}$ to $\sim10^{-3}$. This graph shows that, for the set of $m_N$ values taken into account, CMEs as large as those of the LHC yield contributions to $|\lambda_3^{\rm NP}|$ that can be ${\cal O}\big(10^{-3}\big)$. The same can be said about future hadron colliders, though note that in such a context values $\sim10^{-4}$ seem to be favored. These new-physics contributions are, in the most optimistic case, about one order of magnitude below current experimental sensitivity, as shown in Eq.~\eqref{l3globalfit}. Nonetheless, from Eq.~\eqref{l3HLLHC}, the expected sensitivity of the HL-LHC to the  AC $\lambda_2$ is estimated to be of order $10^{-3}$, which lies close to the largest values of the AC contribution $|\lambda_3^{\rm NP}|$, estimated in the present work. Regarding the $|\lambda_4^{\rm NP}|$ contribution, the graph in the lower panel of Fig.~\ref{l34Sc1} shows that values as large as $\sim10^{-4}$ are achieved. In spite of this, a contribution of order $\sim10^{-5}$ is clearly favored. 
\\ 

\begin{figure}[ht]
\center
\includegraphics[width=7.5cm]{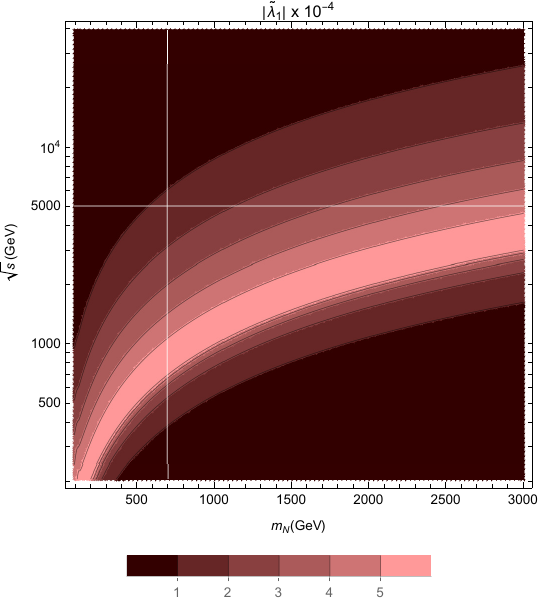}
\vspace{0.3cm}
\\
\includegraphics[width=7.5cm]{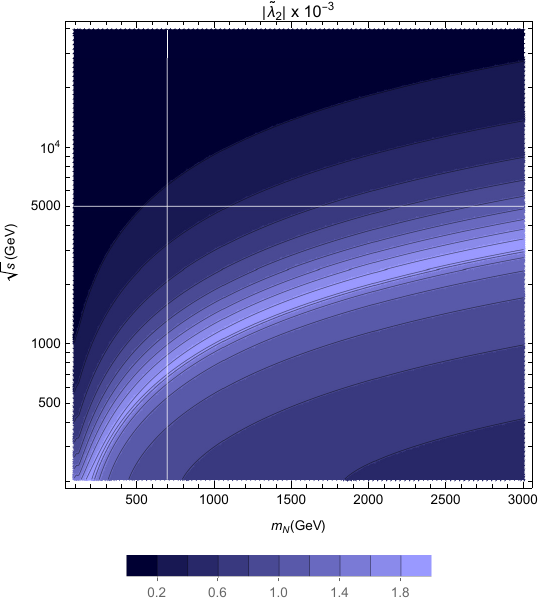}
\caption{\label{tildel12Sc1} Contributions, in the context of $Wh$ production, to $CP$-odd AC $|\tilde{\lambda}_1^{\rm NP}|$ and $|\tilde{\lambda}_2^{\rm NP}|$, plotted in the $\big( m_N,\sqrt{s} \big)$ plane, for fixed $\hat{\rho}=0.65$. Values ranging within $100\,{\rm GeV}\leqslant m_N\leqslant3\,{\rm TeV}$ and $m_h+m_W\leqslant\sqrt{s}\leqslant39\,{\rm TeV}$ have been used. Contributions as large as $|\tilde{\lambda}_1^{\rm NP}|\sim10^{-4}$ and $|\tilde{\lambda}_2^{\rm NP}|\sim10^{-3}$ are generated.}
\end{figure}

For the quantitative discussion on $CP$-violating effects, we refer the reader to the graphs shown in Fig.~\ref{tildel12Sc1},
where the upper panel displays the contributions to $\tilde{\lambda}_1^{\rm NP}$, whereas the $\tilde{\lambda}_2^{\rm NP}$ contributions are represented by the lower graph. As we did in the case of the $CP$-even contributions, the moduli $|\tilde{\lambda}_1^{\rm NP}|$ and $|\tilde{\lambda}_2^{\rm NP}|$ have been estimated. As before, the graphs have been made in the parameter space $\big( m_N,\sqrt{s} \big)$, considering the intervals of values $100\,{\rm GeV}\leqslant m_N\leqslant3\,{\rm TeV}$ and $m_h+m_W\leqslant\sqrt{s}\leqslant39\,{\rm TeV}$. Always keep in mind that $\sqrt{s}$ axes are labeled in base-10 logarithmic scale. Labeling bars, beneath each graph, provide the size of the AC contributions, represented by colored regions, which characterize larger contributions as far as the tone is lighter, whereas darker tones refer to smaller contributions. The graph situated in the upper panel of Fig.~\ref{tildel12Sc1}, portraying the set of $|\tilde{\lambda}_1^{\rm NP}|$ contributions, shows that this AC contribution is typically of order $10^{-4}$. The contributions to $|\tilde{\lambda}_2^{\rm NP}|$, on the other hand, range from $\sim10^{-4}$ to $10^{-3}$. By comparing the $|\tilde{\lambda}_1^{\rm NP}|$ contribution to the results of Ref.~\cite{ATLAS:2019jst}, displayed in Eq~\eqref{l1ATLAS}, we conclude that these $CP$-violating effects lie far away from current experimental sensitivity,  by about 3 orders of magnitude. The expected improvement from CLIC, Eq.~\eqref{l1CLIC}, shortens this difference to $\sim2$ orders of magnitude.
\\


\subsection{Scenario 2: \texorpdfstring{$W^{*}W^{*}h$}{W*W*h}}
\begin{figure}[ht]
\center
\includegraphics[width=6cm]{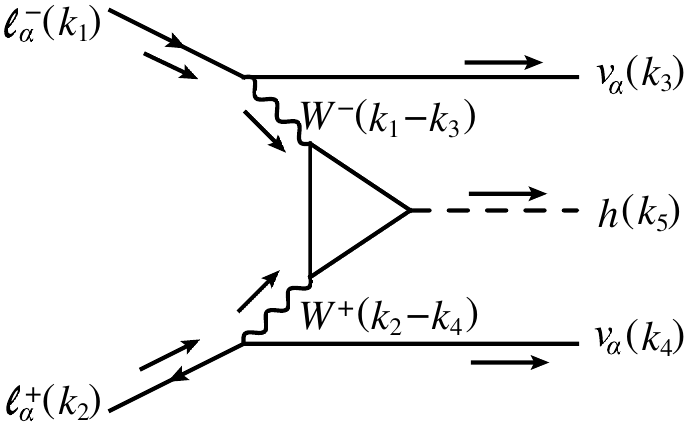}
\caption{\label{wwhindiags}Generic Feynman diagram illustrating how light and heavy neutrinos contribute to Higgs production by VBF, in the context of some lepton collider.}
\end{figure}

To address this possibility, we consider Higgs production by VBF in some lepton collider, which, as illustrated by the diagram of Fig.~\ref{wwhindiags}, 
receives contributions from the $WWh$ vertex. In contrast with the case presented in the previous Section, the occurrence of 2 virtual lines prevents the CME of the collision from directly determining the allowed values for the off-shell $W$-boson squared 4-momenta, $(k_1-k_3)^2$ and $(k_2-k_4)^2$. In order to have some reference about what values to take, we note that the cross section for the process $\ell\ell\to\nu\nu h$ can be written as
\begin{equation}\label{sigmasc2}
  \sigma= \frac{1}{2 (2\pi)^5 \lambda^{1/2}(s,m^2_\ell,m^2_\ell)}\int d \Pi_3 \big| \mathcal{M}\big|^2,
\end{equation}
where, $\lambda^{1/2}(s,m^2_\ell,m^2_\ell)$ is the K\"allen lambda function~\cite{Kallen:1964lxa}, which depends on the initial-lepton masses and the CME, $\sqrt{s}$, in the global center-of-mass (CM) frame. From here on, we refer to the global CM frame by $M_{\rm CM}$. Moreover, $d \Pi_3$ is the three-body Lorentz invariant phase space, whereas the amplitude for the process $\ell\ell\to\nu\nu h$ is denoted by $\mathcal{M}$. To evaluate this integral, it is convenient to introduce an intermediate state $q$ that decays into the pair $\nu_{\alpha} + \nu_{\alpha}$. This intermediate state can be represented as:
\begin{equation}
    \ell^{-}_{\alpha} +  \ell^{+}_{\alpha} \to q + h \to (\nu_{\alpha} + \nu_{\alpha} )+ h. 
\end{equation}
Using this intermediate state $q$, we can express the phase-space integral as the product of two processes: a $2 \to 2$ scattering followed by a $1\to 2$ decay~\cite{Byckling:1971vca}, that is,
\begin{equation}
    \begin{aligned}
    \int d \Pi_3 \big| \mathcal{M}\big|^2 = \int d m_{34} d\Pi_{2}(s,m^2_h,m^2_{34})d\Pi_{2}(m^4_{34},m^2_{\nu},m^2_{\nu}),
    \end{aligned}
\end{equation}
where $d\Pi_{2}$ refers to the two-body phase-space element. Since the phase-space integral is Lorentz invariant, it can be evaluated in any reference frame. With this in mind, for the scattering process $\ell^{-}_{\alpha} + \ell^{+}_{\alpha} \to q + h$ we use the $M_{\rm CM}$ frame, whereas for the subsequent $2\to1$ process the CM frame for the pair $\nu_{\alpha} + \nu_{\alpha}$, in which $\vec{k}_3+\vec{k}_4=0$ holds, is chosen. In what follows, this reference frame is denoted by $M_{\rm CM}^*$.
 Then, the cross section can be expressed as
\begin{widetext}
\begin{equation}\label{eq:1}
    \begin{aligned}
        \sigma = \frac{1}{64 s (2\pi)^4 \lambda^{1/2}(s,m^2_\ell,m^2_\ell)}\int^{m^{\textrm{max}}_{34}}_{m^{\textrm{min}}_{34}}\frac{dm_{34}}{m_{34}}  \lambda^{1/2}(s,m^2_h,m^2_{34})\lambda^{1/2}(m^2_{34},m^2_\nu,m^2_\nu) \int^{\pi}_{0}d\theta\,\sin\theta \int d\Omega^{*} \big| \mathcal{M}\big|^2.
    \end{aligned}
\end{equation}
\end{widetext}
In this equation, the polar angle $\theta$ is associated with the Higgs boson in $M_{\rm CM}$, while $(\theta^*, \varphi^*)$ describes the direction of one of the final-state neutrinos $\nu_{\alpha}$, with respect to the direction of motion of the pair $\nu_{\alpha} + \nu_{\alpha}$ in $M_{\rm CM}^*$. Keep in mind that all quantities with an asterisk are given with respect to the $M_{\rm CM}^*$ frame, while those without an asterisk refer to the $M_{\rm CM}$ frame. Finally, $m^{2}_{34}$ is the invariant mass of the pair $\nu_{\alpha} + \nu_{\alpha}$. If we neglect the masses of the initial-state leptons, as well as the masses of the final-state neutrinos, the relevant kinematic quantities to our work are  
\begin{eqnarray}
\label{eq:2}
& \displaystyle E_h = \frac{s + m^2_h-m^{2}_{34}}{2 \sqrt{s}}, & \hspace{0.3cm} E^{*}_{\nu} = \frac{m_{34}}{2},
\\
\label{eq:2a}
& \displaystyle E_{34} = \frac{s + m^{2}_{34}-m^2_h}{2 \sqrt{s}}, & \hspace{0.3cm} | \vec{k}^{*}_3| = \frac{m_{34}}{2}.
\end{eqnarray}
In addition, all physical quantities depend on $m_{34}$, whose minimum and maximum values (shown in Eq.~\eqref{eq:1}) are given by
\begin{equation}
    0 \leq m^2_{34} \leq \left( \sqrt{s}-m_h \right)^2 . 
\end{equation}
We are interested in
\begin{eqnarray}
\label{eq:3}
&& k^{2}_{13} = \left(k_1-k_3 \right)^2 = -2 k_1 \cdot k_3,
\\ \nonumber \\
\label{eq:4}
&& k^{2}_{24} = \left(k_2-k_4 \right)^2 = -2 k_2 \cdot k_4,
\end{eqnarray}
which correspond to the squared 4-momentum transfers of the virtual $W^{\mp}$ bosons, as displayed in Fig.~\ref{wwhindiags}. In Eqs.~(\ref{eq:3})-(\ref{eq:4}), the 4-vectors $k_3$ and $k_4$ are given in $M_{\rm CM}$, though, aiming at the resolution of the cross section, these 4-vectors are calculated in $M_{\rm CM}^*$. The two reference frames are connected by the following Lorentz transformation:
\begin{equation}
    k_j = \textrm{Rot}_{y}\left( \theta+\pi \right)\textrm{Boost}_{z} \left( \beta \right)k^{*}_j,
\end{equation}
with $j=3,4$ and where
\begin{equation}\label{eq:5}
\begin{aligned}
    \beta = \sqrt{1-\frac{1}{\gamma^2}}, && \gamma = \frac{E_{34}}{m_{34}}. 
\end{aligned}
\end{equation}
Note that the energy $E_{34}$ is given in Eq.~\eqref{eq:2a}. An energy range for the squared 4-momenta of the virtual $W^{\mp}$ bosons can be established using the equations~\eqref{eq:2} to~\eqref{eq:5}, from which we conclude that 
\begin{eqnarray}
&&\displaystyle -s \left(1-\frac{m^2_h}{s} \right)< k^2_{13} \leq 0, 
\label{k13squared}
\\
&&\displaystyle -s \left(1-\frac{m^2_h}{s} \right)< k^2_{24} \leq 0.
\label{k23squared}
\end{eqnarray}
\\

\begin{figure}[ht]
\center
\includegraphics[width=7.5cm]{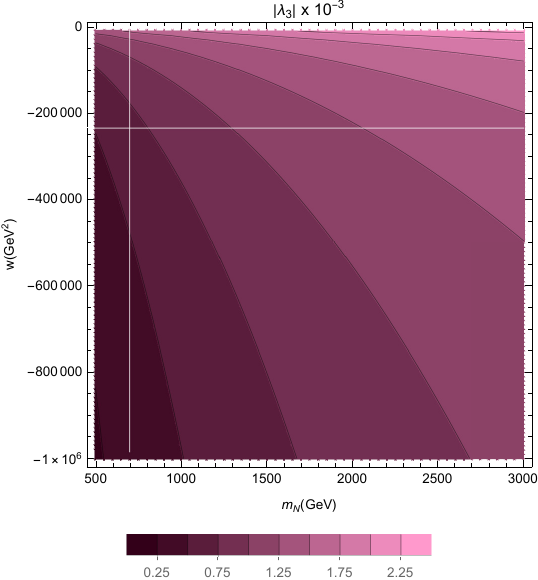}
\vspace{0.3cm}
\\
\includegraphics[width=7.5cm]{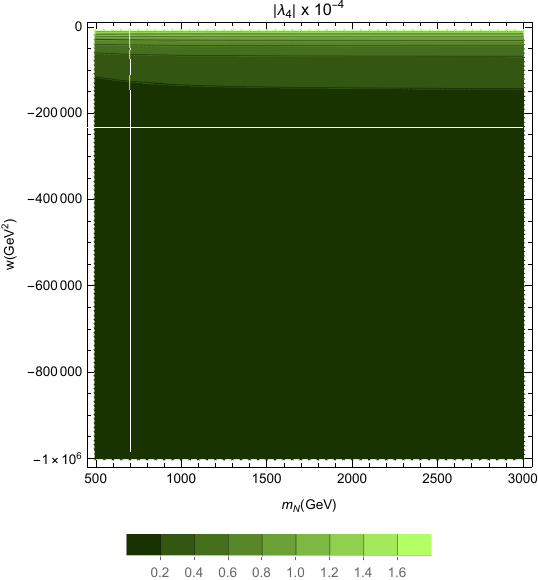}
\caption{\label{CPevenSc2} Contributions, in the context of Higgs production by VBF, to $CP$-even AC $|\lambda_3^{\rm NP}|$ and $|\lambda_4^{\rm NP}|$, plotted in the $\big( m_N,w \big)$ plane, for fixed $\hat{\rho}=0.65$. Values ranging within $500\,{\rm GeV}\leqslant m_N\leqslant3\,{\rm TeV}$ and $-1\,{\rm TeV}^2\leqslant w\leqslant0\,{\rm TeV}^2$ have been used. Contributions as large as $|\lambda_3^{\rm NP}|\sim10^{-3}$ and $|\lambda_4^{\rm NP}|\sim10^{-4}$ are generated.}
\end{figure}

For our discussion on the $CP$-preserving effects, consider Fig.~\ref{CPevenSc2},
which shows graphs for the moduli $|\lambda_3^{\rm NP}|$ and $|\lambda_4^{\rm NP}|$. Recall that, in this scenario, not one but two squared momenta of virtual particles are to be varied, namely, $k_{13}^2$ and $k_{24}^2$. To effectuate the graphs of Fig.~\ref{CPevenSc2}, we have taken such quantities to be equal and then we have denoted $w\equiv k_{13}^2=k_{24}^2$. Under such circumstances, the graphs have been plotted in the $\big( m_N,w \big)$ parameter space, where $500\,{\rm GeV}\leqslant m_N\leqslant3\,{\rm TeV}$ and $-1\,{\rm TeV}^2\leqslant w\leqslant0\,{\rm GeV}^2$ have been considered. As before, the colors characterizing each region correspond to different values of the muduli $|\lambda_3^{\rm NP}|$ or $|\lambda_4^{\rm NP}|$, with the lighter tones representing the larger contributions. The vertical lines, appearing in both graphs, represent the minimal allowed heavy-neutrino mass, in accordance with Ref.~\cite{CMS:2018iaf}, by the CMS Collaboration. The horizontal line, on the other hand, is the minimal $w$ value for an lepton collider with a CME of $\sqrt{s}=500\,{\rm GeV}$, which we have determined by usage of Eqs.~\eqref{k13squared} and \eqref{k23squared}. We observe that contributions to the $CP$-even anomalous form factor $|\lambda_3^{\rm NP}|$ can be as large as $10^{-3}$, whereas $|\tilde{\lambda}_4^{\rm NP}|$ reaches values as large as $\sim10^{-4}$. These results are in agreement with our results for hadron colliders, previously discussed and, again, suggest that the new-physics AC contribution $\lambda_3^{\rm NP}$ could lie close to HL-LHC sensitivity.
\\ 

\begin{figure}[ht]
\center
\includegraphics[width=7.5cm]{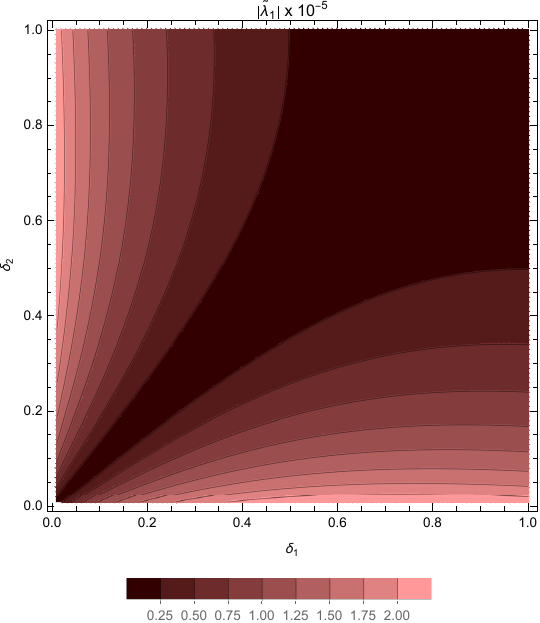}
\vspace{0.3cm}
\\
\includegraphics[width=7.5cm]{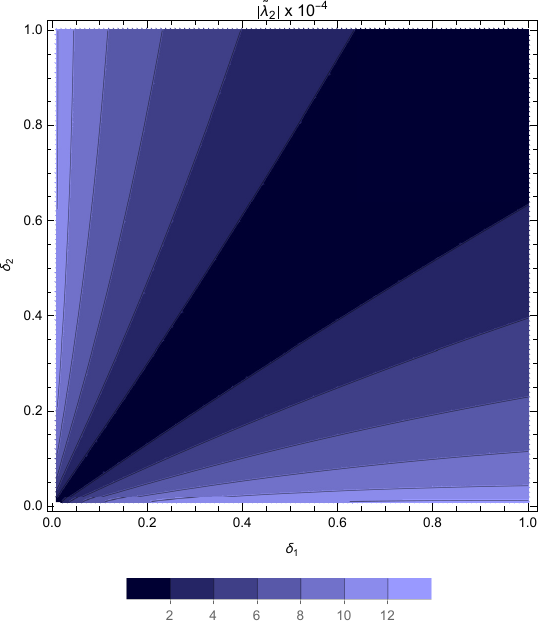}
\caption{\label{CPoddSc2primeras} Contributions, in the context of Higgs production by VBF, to $CP$-odd AC $|\tilde{\lambda}_1^{\rm NP}|$ and $|\tilde{\lambda}_2^{\rm NP}|$, plotted in the $\big( \delta_1,\delta_2 \big)$ plane, for fixed $\hat{\rho}=0.65$, $m_N=700\,{\rm GeV}$, and $w=-\big( 484\,{\rm GeV} \big)^2$. Values ranging within $0\leqslant \delta_1\leqslant1$ and $0\leqslant \delta_2\leqslant1$ have been used. Contributions as large as $|\tilde{\lambda}_1^{\rm NP}|\sim10^{-5}$ and $|\tilde{\lambda}_2^{\rm NP}|\sim10^{-4}$ are generated.}
\end{figure}

Now we discuss the $CP$-violating effects. We start noting that once the heavy-neutrino mass spectrum is taken degenerate and the matrix texture given in Eq.~\eqref{xitexture} is implemented, the $CP$-odd contributions can be expressed as
\begin{equation}
\tilde{\lambda}_j^{\rm NP}=\Omega-2i\sin2\phi\,\Big( \frac{\hat\rho}{1+\hat\rho} \Big)^2\sum_\alpha\sum_{j=1}^3|U_{\rm PMNS}|\,\eta_\alpha^{\nu N},
\label{CPoddlambdastructure}
\end{equation}
with $\Omega$ and $\eta_\alpha^{\nu N}$ some functions of masses. Let us remark that the second term of this equation comprises the whole dependence on the phase $\phi$, and clearly vanishes for $\phi=0,\frac{\pi}{2}$. However, it is the $\Omega$ term the one which dominates the contribution, being larger that the $\phi$-term by about 11 orders of magnitude. Therefore, the choice of the phase is not really important in practical terms. We have noticed that taking $k_{13}^2=k_{24}^2$ yields an exact elimination of the $\Omega$-term, that is, $\Omega=0$, thus severely suppressing the contributions. Then the two $CP$-violating contributions $\tilde{\lambda}_1^{\rm NP}$ and $\tilde{\lambda_2^{\rm NP}}$ vanish if the momenta of the two virtual $W$ bosons are taken to be equal and the phase $\phi$, defined in Eq.~(\ref{xitexture}), is taken as $\phi=0$. Evidently, to determine how large $\tilde{\lambda}_1^{\rm NP}$ and $\tilde{\lambda}_2^{\rm NP}$ can be, $k_{13}^2\ne k_{24}^2$ must be considered for the discussion. In what follows, we write the virtual $W$-boson squared momenta as $k_{13}^2=\delta_1 w$ and $k_{24}^2=\delta_2 w$, in order to explore the framework in which $k_{13}^2\ne k_{24}^2$. Here,
\begin{equation}
-s\Big( 1-\frac{m_h^2}{s} \Big)\leqslant w\leqslant 0\,{\rm GeV},
\end{equation}
\begin{equation}
0\leqslant \delta_1\leqslant 1,
\label{delta1}
\end{equation}
\begin{equation}
0\leqslant \delta_2\leqslant 1.
\label{delta2}
\end{equation}
The behavior of the moduli $|\tilde{\lambda}_1^{\rm NP}|$ and $|\tilde{\lambda}_2^{\rm NP}|$ with respect to $\delta_1$ and $\delta_2$ is illustrated by the graphs of Fig.~\ref{CPoddSc2primeras},
which have been plotted for the fixed value $m_N=700\,{\rm GeV}$, of the heavy-neutrino mass, and $w=-(484\,{\rm GeV})^2$, also fixed, while the factors $\delta_1$ and $\delta_2$ have been varied in accordance with Eqs~\eqref{delta1} and \eqref{delta2}. As expected from our discussion on the analytical structure of the $CP$-violating AC contributions, Eq.~\eqref{CPoddlambdastructure}, we note that the $\tilde{\lambda}_j^{\rm NP}$ values associated to $\delta_1=\delta_2$ fall into the darkest regions in each graph. Such contributions are quite small, of order $10^{-12}$, in contrast with the contributions corresponding to the lightest tones in the graphs, which are as large as $~10^{-5}$, in the case of $|\tilde{\lambda}_1^{\rm NP}|$, and $10^{-4}$ for $|\tilde{\lambda}_2^{\rm NP}|$. These graphs show that, as long as the difference among $k_{13}^2$ and $k_{24}^2$ is large, the AC contributions $|\tilde{\lambda}_1^{\rm NP}|$ and $|\tilde{\lambda}_2^{\rm NP}|$ acquire their most significant values.
\\

\begin{figure}
\center
\includegraphics[width=8cm]{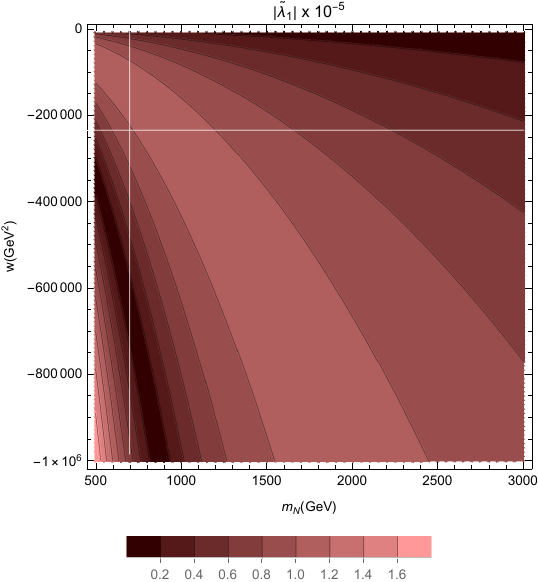}
\vspace{0.3cm}
\\
\includegraphics[width=8cm]{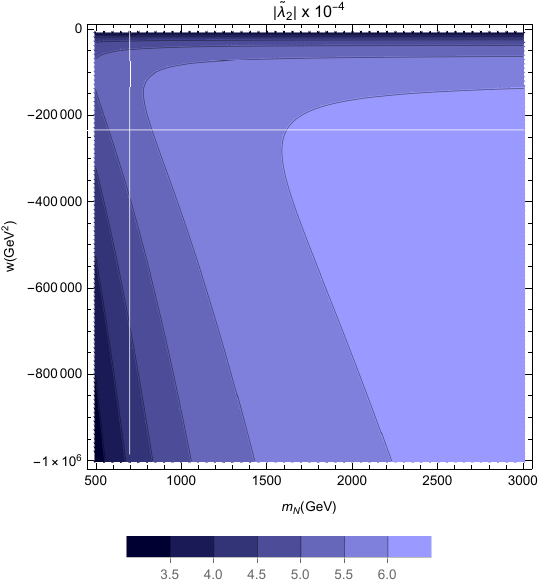}
\caption{\label{CPoddSc2} Contributions, in the context of Higgs production by VBF, to $CP$-odd AC $|\tilde{\lambda}_1^{\rm NP}|$ and $|\tilde{\lambda}_2^{\rm NP}|$, plotted in the $\big( m_N,w \big)$ plane, for fixed $\hat{\rho}=0.65$. Values ranging within $500\,{\rm GeV}\leqslant m_N\leqslant3\,{\rm TeV}$ and $-1\,{\rm TeV}^2\leqslant w\leqslant100\,{\rm GeV}^2$ have been used. Contributions as large as $|\tilde{\lambda}_1^{\rm NP}|\sim10^{-5}$ and $|\tilde{\lambda}_2^{\rm NP}|\sim10^{-4}$ are generated.}
\end{figure}

With the previous discussion in mind, we consider the values $\delta_1=\frac{1}{5}$, $\delta_2=\frac{2}{3}$, and then we estimate the contributions $|\tilde{\lambda}_1^{\rm NP}|$ and $|\tilde{\lambda}_2^{\rm NP}|$, which is illustrated by the graphs in Fig.~\ref{CPoddSc2}.
The graphs displayed in this figure have been carried out in the $\big( m_N,w \big)$ parameter space, with the parameters ranging within $500\,{\rm GeV}\leqslant m_N\leqslant3\,{\rm TeV}$ and $-1\,{\rm TeV}^2\leqslant w\leqslant-100\,{\rm GeV}^2$. As before, the colored regions, associated to different values of either $|\tilde{\lambda}_1^{\rm NP}|$ or $|\tilde{\lambda}_2^{\rm NP}|$, correspond to larger values as the tone of the region is lighter, whereas the represented contributions are smaller for darker regions. As anticipated, these graphs show that values as large as $\sim10^{-5}$, for $|\tilde{\lambda}_1^{\rm NP}|$, and $\sim10^{-4}$, for $|\tilde{\lambda}_2^{\rm NP}|$, characterize these $CP$-odd anomalous contributions. 
\\


\section{Summary}
\label{theend}
The aim of this work has been the calculation, estimation and discussion of contributions, at the one-loop level, from Majorana neutrinos, heavy and light ones, to the Higgs gauge coupling $WWh$, in the context of a seesaw variant in which the type-1 seesaw link among the masses of heavy and light neutrinos is broken by the elimination of tree-level light-neutrino masses, which are then generated radiatively, thus defining a new connection among such masses, in which the tininess of light neutrinos is achieved if the heavy-neutrino mass spectrum is near-degenerate. Part of the structure of the resulting $WWh$ vertex function was found to match the one generated by the mass-dimension-6 SM effective-Lagrangian terms in the so-called SILH basis, after implementation of electroweak symmetry breaking. This parametrization involves a $CP$-even coupling accompanying the Lorentz-covariant structure of the SM tree-level $WWh$ vertex. Two $CP$-preserving AC, here denoted by $\lambda_2^{\rm NP}$ and $\lambda_3^{\rm NP}$, are also present in this vertex-function structure. Since our interest focused in the AC, we have disregarded the SM-like coupling. Moreover, our explicit one-loop calculation showed that $\lambda_2^{\rm NP}=0$, so the only nonzero SILH-type AC was $\lambda_3^{\rm NP}$. The sole $CP$-violating AC generated by the SILH Lagrangian, here referred to by $\tilde{\lambda}_1$, was also present in the one-loop $WWh$ vertex function. Further couplings, associated to effective-Lagrangian terms of mass dimension $>6$, were also found in this vertex function. Those associated to $CP$ invariance were denoted by $\tilde{\lambda}_4^{\rm NP}$ and $\tilde{\lambda}_5^{\rm NP}$, whereas a $CP$-odd-related AC, $\tilde{\lambda}_2^{\rm NP}$, also emerged. According to our calculation, the relation $\lambda_4^{\rm NP}=\lambda_5^{\rm NP}$ holds. Therefore, in total, the calculation yielded 2 $CP$-even and 2 $CP$-odd AC. All the AC comprised by the vertex function were found to be free of UV divergences. In order to get an estimation of the size of the AC contributions, we considered two scenarios. In one of such scenarios, one of the $W$-boson lines was taken off the mass shell, with the other $W$-boson line and the Higgs-boson line assumed to be on shell. This possibility was contextualized by considering the vertex $WWh$ as a piece of a subprocess $ud\to Wh$, occurring in $pp$ collisions. This allowed us to estimate the $CP$-even and $CP$-odd contributions to anomalous gauge Higgs couplings. In the case of the former, we found contributions as large as $\sim10^{-3}$, which are near the expected sensitivity of the HL-LHC. As far as $CP$-violation was concerned, the contributions were found to be below the expected sensitivity for CLIC by about 2 orders of magnitude. Another framework considered for $WWh$ was defined by taking the pair of $W$-boson lines off shell, but the Higgs-boson line on shell. To provide a framework for the estimations and discussion, we noted that $WWh$ contributes to the process of Higgs production by means of VBF, $\ell\ell\to\nu\nu h$, which we considered in the context of a lepton collider. Among the free parameters, there were the squared momenta of the virtual $W$ bosons, which we varied taking as our reference the restrictions imposed by the phase space of the corresponding cross section. The largest $CP$-even contributions were found to be of order $10^{-3}$, for $|\lambda_3^{\rm NP}|$, and $10^{-4}$, in the case of $|\lambda_4^{\rm NP}|$.


\section*{Acknowledgements}
\noindent
We acknowledge financial support from Secihti (M\'exico). E. R and M.S. are grateful for the financial support provided by the program \textit{Estancias Posdoctorales por México}. H. V. thanks the funding through the program ``\textit{Becas Nacionales para estudios de Posgrado 2024-2}", with support number 4037802.



\bibliographystyle{apsrev4-1}
\bibliography{HNHWW}

\end{document}